\documentclass[twocolumn, prx, reprint, amsfonts, amsmath, amssymb, superscriptaddress, longbibliography]{revtex4-2}
\usepackage{graphicx}
\usepackage[colorlinks,linkcolor=blue,anchorcolor=red,citecolor=green]{hyperref}

\begin{document}


\title{The Berry dipole photovoltaic demon and the thermodynamics of  photo-current generation within the optical gap of metals}

\author{Li-kun Shi}
\affiliation{Max-Planck-Institut für Physik komplexer Systeme, Nöthnitzer Straße 38, 01187 Dresden, Germany}

\author{Oles Matsyshyn}
\affiliation{Division of Physics and Applied Physics, Nanyang Technological University, Singapore 637371, Republic of Singapore}
\affiliation{Max-Planck-Institut für Physik komplexer Systeme, Nöthnitzer Straße 38, 01187 Dresden, Germany}

\author{Justin C. W. Song}
\affiliation{Division of Physics and Applied Physics, Nanyang Technological University, Singapore 637371, Republic of Singapore}

\author{Inti Sodemann Villadiego}
\affiliation{Institut für Theoretische Physik, Universität Leipzig, Brüderstraße 16, 04103, Leipzig, Germany}
\affiliation{Max-Planck-Institut für Physik komplexer Systeme, Nöthnitzer Straße 38, 01187 Dresden, Germany}
\date{\today}

\begin{abstract}
We dismantle the previously held misconception that it is impossible for bulk rectification mechanisms to induce a net DC electric current when the frequency of the impinging radiation lies within the optical gap of a metal in the limit of small carrier relaxation rates. We argue that generically such in-gap rectification mechanisms are irreversible and accompanied by a continuous exchange of energy with a heat bath and must also be necessarily accompanied by a small but finite absorption of radiation in order to guarantee the positivity of the net entropy production and abide by the second law of thermodynamics. We show, however, that the intra-band non-linear Hall effect arising from the Berry curvature is a special kind of in-gap rectification mechanism that behaves as a ``photo-voltaic demon'', namely it can operate as an ideal reversible and dissipationless conveyor of energy between the radiation and an external circuit. Its reversible nature allows for an interesting mode of operation as an amplifier of circularly polarized light, whose efficiency can approach 100\%, and which could be technologically promising especially in the infrared frequency range.
\end{abstract}




\maketitle

\section*{Introduction}
Materials with broken inversion symmetry can display {\it bulk rectification effects}, whereby an oscillating electric field produces an average rectified DC electric current. While these effects have been investigated for decades~\cite{kraut1979anomalous,belinicher1980photogalvanic,von1981theory,belinicher1982kinetic}, there is a recent upsurge of interest in investigating their interplay with the electronic band structure and Berry phase geometry~\cite{aversa1995nonlinear,sipe2000second,moore2010confinement,young2012first,sodemann2015quantum,morimoto2016topological,chan2017photocurrents,de2017quantized,nagaosa2017concept,vanderbilt2018berry,kang2019nonlinear,ma2019observation,matsyshyn2019nonlinear,parker2019diagrammatic,sturman2020ballistic,matsyshyn2021berry,shi2021geometric,xu2021pure,wang2022generalized}, as well as their potential for novel opto-electronic technologies~\cite{young2012first,matsyshyn2021berry,brehm2014first,rangel2017large,cook2017design,morimoto2018current,kumar2021room}.

Despite all this research activity, the understanding of how these bulk rectification effects fit within the conceptual framework of non-equilibrium thermodynamics is relatively un-explored. Therefore, the first major objective of our current study is to contribute to fill in this gap by investigating perturbatively the constraints imposed by the second law of thermodynamics on the leading non-linear response functions that govern such bulk rectification effects. 

A second major objective of our work is to demonstrate that it is possible to have a finite DC rectified current when the frequency of radiation lies within the optical gap of a material in the clean limit of small carrier relaxation rates. While examples of in-gap rectification have been discussed recently~\cite{kaplan2020nonvanishing,gao2021intrinsic,watanabe2021chiral}, earlier prominent studies~\cite{belinicher1986transient,ivchenko1988magneto} had concluded that such current rectification within the optical gap of a material was impossible in the clean limit. Hence, our second major objective is basically to try to dismantle this fundamental misconception that has been spread by past work~\cite{belinicher1986transient,ivchenko1988magneto}, namely, we will demonstrate that there can be a net rectified DC current when the frequency of the driving oscillating electric field lies within the optical gap of the electronic band structure in the ideal limit of zero temperature and vanishingly small relaxation rates and to second order in the driving electric fields. By using a microscopically explicit model of an electronic system coupled to a heat bath, we will show that this possible for metallic systems with a Fermi surface and we will also discuss why this is consistent with the laws of thermodynamics.

From the existence of these in-gap rectification effects, one might be tempted to conclude that such mechanisms could induce DC electric photo-currents without an accompanying light absorption. As, we will see however, despite remaining finite in the limit of small relaxation rates, such in-gap rectification mechanisms are generically {\it dissipative}, in the sense that they are generically accompanied by a net positive entropy production. As a consequence, they are generically accompanied by a small but finite photon absorption, which must be present in order to abide by the second law of thermodynamics. Therefore, it is inaccurate to claim that these in-gap rectification mechanisms are not accompanied by photon absorption, as recently stated in Ref.~\cite{onishi2022photovoltaic}. We have found however one special limit in which one particular in-gap rectification mechanism behaves as a {\it non-dissipative} reversible mechanism that does not contribute to the net entropy production and, ideally, does not need to be accompanied by irreversible light absorption. This mechanism is the non-linear Hall effect~\cite{moore2010confinement,sodemann2015quantum,kang2019nonlinear,ma2019observation,matsyshyn2019nonlinear,deyo2009semiclassical}.

As we will show, in the limit of frequencies smaller compared to the optical gap but larger than the relaxation rate, the ``Hall'' nature of this non-linear Hall effect allows to transfer the energy of circularly polarized light onto the energy of an external electric circuit and vice-versa in a reversible non-dissipative fashion, which is why we refer this mechanism ``as a photovoltaic demon''. The third major objective of our study will be then to illustrate the interesting opportunities that this mechanism offers for novel opto-electronic technologies. More specifically, we will show that the efficiency of this mechanism to convert the energy of circularly polarized light onto DC electric energy can approach 100\% in the limit of  small frequencies compared to the optical gap. But perhaps, more interestingly, because of its reversibility, the same mechanism can be used to transfer energy from at DC circuit onto the radiation with high efficiency and therefore act as an effective amplifier of low frequency circularly polarized light.

Our paper is organized as follows. In Section~\ref{Thermodynamics}, we setup the framework that incorporates both thermodynamics and nonlinear responses, discuss the work performed by the radiation and the circuit, and identify the key quantities that allow to determine whether an in-gap rectification mechanism is dissipative or not. In Section~\ref{BoltzmannDescription} we illustrate these principles and quantities within a simplified Boltzmann single-band description. Section~\ref{QuantumFormalism} discusses a microscopic description of the bulk rectification in the presence of a physical heat bath. Section~\ref{QuantumResults} applies the general considerations of Sections~\ref{Thermodynamics} and~\ref{QuantumFormalism} to a specific model, and validates the simpler picture of Section~\ref{BoltzmannDescription}. In Section~\ref{Applications}, we discuss  photovoltaic and light-amplification devices based on these principles, their efficiency and the requirements for their operation.

\section{Thermodynamic considerations}
\label{Thermodynamics}
We consider a crystalline electronic system coupled to a heat bath and subjected to a spatially uniform but time dependent vector potential ${\bf A}(t)$. The energy of the system can change in two ways: by the {\it work}, $\Delta W$, performed by the vector potential ${\bf A}(t)$, and by the {\it heat}, $\Delta Q$, absorbed or released into the bath. From the density matrix describing the system, $\rho_S(t)$, these two quantities can be computed as follows~\cite{jarzynski1997nonequilibrium,landau2013statistical}:

\begin{align}
&\Delta  W = \int_{t_i}^{t_f} \text{d} t \ \text{tr}\left(\rho_S \frac{\text{d} H_S}{\text{d} t}\right) = \int_{t_i}^{t_f} \text{d} t \ {\bf j}(t)\cdot {\bf E}(t),\label{DeltaW}\\
&\Delta  Q = \int_{t_i}^{t_f} \text{d} t \ \text{tr}\left( H_S \frac{\text{d} \rho_S}{\text{d} t}\right),
\end{align}
where $t_i$ $t_f$ are initial and final times of a process and $H_S$ is the Hamiltonian of the electronic system. The second equality of the expression for the work can be obtained by assuming that the only explicit time dependent parameter changing in the Hamiltonian of the system is the vector potential ${\bf A}(t)$. The above makes manifest that the change of energy of the system is $\Delta E=\Delta W+\Delta Q$.

\begin{figure}
\includegraphics[width=0.49\textwidth]{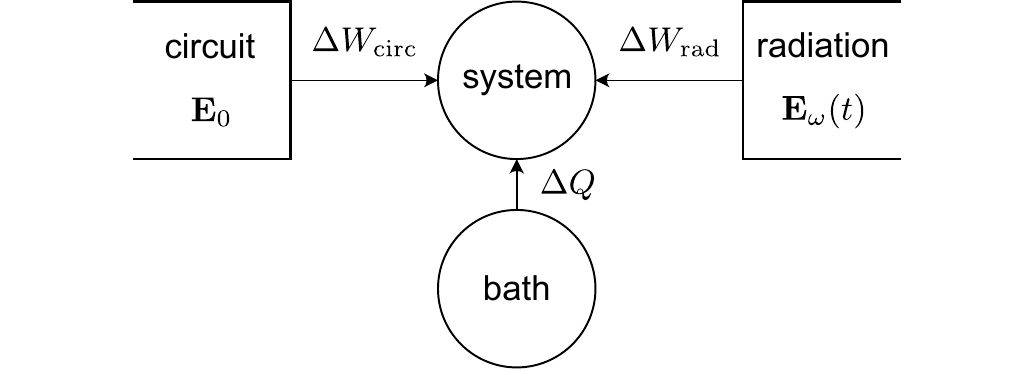}
\caption{Schematic of the crystalline electronic system coupled to a heat bath, connected to an external circuit and subject to a radiation field.
}
\label{fig-schematic}
\end{figure}

We would like to investigate the energy exchange of the system with the radiation and an external electric circuit, as depicted in Fig.~\ref{fig-schematic}. We view the external electric circuit as providing a DC time independent electric field, ${\bf E}_0$, and the radiation as the source of an oscillating electric field with frequency $\omega$, ${\bf E}_\omega (t) = {\bf E}_\omega e^{i \omega t} + \text{c. c.}$, where ${\bf E}_\omega$ is vector that can be complex to account for the degree of polarization of light. The total electric field acting on the system is the sum of these ${\bf E}(t)= {\bf E}_0 + {\bf E}_\omega (t)$, and therefore, from Eq.~(\ref{DeltaW}), the work can be partitioned into the work performed by the circuit and the radiation $\Delta  W = \Delta W_\text{circ} + \Delta W_\text{rad}$, where
\begin{align}
\Delta W_\text{circ} = \int_{t_i}^{t_f} \text{d} t \ {\bf j}(t)\cdot {\bf E}_0,
\
\Delta W_\text{rad} = \int_{t_i}^{t_f} \text{d} t \ {\bf j}(t)\cdot {\bf E}_\omega (t).
\label{DeltaW-partitioned}
\end{align}
Let us assume the system reaches a well defined steady state of oscillations periodic in the drive, with period $T = 2\pi/\omega$. Because of periodicity the change of the system energy vanishes over one cycle: $\Delta E = 0$~\footnote{Due to the DC electric field, strictly speaking the Hamiltonian is not periodic in time, but it is periodic up to a gauge transformation after one period}. On the other hand the Kelvin-Planck statement of the second law of thermodynamics~\cite{zemansky1997heat}, implies that during one cycle the system can only release heat into the bath: $\Delta Q \leq 0$. Therefore the second law of thermodynamics implies that the net work performed on the system must be non-negative:
\begin{align}
\Delta  W = \Delta W_\text{circ} + \Delta W_\text{rad} 
\geq 0 .
\label{DeltaW-non-negative}
\end{align}
We can compute the above work to leading order in electric fields from linear response theory, where the electric current is given by:
\begin{align} 
{\bf j}^{(1)} (t) = {\bf j}_0^{(1)} + \left( {\bf j}_\omega^{(1)} e^{i \omega t}+ \text{c.c.} \right),
\quad
{\bf j}_\omega^{(1)} = \boldsymbol{\sigma} (\omega) {\bf E}_\omega .
\end{align}
Here ${\boldsymbol \sigma} (\omega)$ is the complex linear conductivity tensor. By inserting the above expression onto Eq.~(\ref{DeltaW-partitioned}), we then obtain the leading expressions for the average power:
\begin{align}
& \frac{ \Delta  W_\text{circ}^{(2)} }{T} = {\bf E}_0^\text{T} {\boldsymbol \sigma} (0) {\bf E}_0^{} , 
\nonumber\\
& \frac{ \Delta  W_\text{rad}^{(2)} }{T} = {\bf E}_\omega^\dagger [ {\boldsymbol \sigma} (\omega) + {\boldsymbol \sigma}^\dagger (\omega) ] {\bf E}_\omega ,
\label{DeltaW-2nd}
\end{align}
where $\Delta  W_\text{circ}^{(2)} / T$ is the Joule heating effect and $\Delta  W_\text{rad}^{(2)} / T$ accounts for the light absorption at finite frequency. Therefore the second law of thermodynamics, as stated in Eq.~(\ref{DeltaW-non-negative}), implies that the symmetric part of the DC conductivity and the Hermitian symmetrized finite frequency conductivity must be non-negative tensors.

Let us now compute the work to the next order of perturbation theory. To second order in fields the current is given by:
\begin{align}
& {\bf j}^{(2)} (t) = {\bf j}_0^{(2)}+ \left( {\bf j}_\omega^{(2)} e^{i\omega t}+{\bf j}_{2\omega}^{(2)} e^{2 i\omega t} + \text{c.c.} \right) ,
\nonumber\\
& {\bf j}_{\omega_f}^{(2)} = \sum_{\omega_1, \omega_2} \delta( \omega_1 +\omega_2 - \omega_f ) \boldsymbol{\sigma}(\omega_1,\omega_2) {\bf E}_{\omega_1} {\bf E}_{\omega_2}
\end{align}
where $\omega_{1,2} \in \{0, \pm \omega \}$, ${\bf E}_{-\omega} \equiv {\bf E}_{\omega}^*$, and $\boldsymbol{\sigma}(\omega_1,\omega_2)$ is the symmetrized second order conductivity tensor, namely 
\begin{align}
\sigma_{abc} (\omega_1,\omega_2) = \sigma_{acb} (\omega_2,\omega_1) .
\label{symmetrized-sigma}
\end{align}

From the above and using Eq.~(\ref{DeltaW-partitioned}) the next order contributions to the circuit and radiation work can be shown to be:
\begin{align}
& \frac{ \Delta  W_\text{circ}^{(3)} }{T} = {\bf E}_0^\text{T} {\boldsymbol \sigma} (0,0) {\bf E}_0^{} {\bf E}_0^{}
+ {\bf E}_0^\text{T} {\boldsymbol \sigma} (\omega,-\omega) {\bf E}_\omega {\bf E}_{\omega}^* , 
\nonumber \\
& \frac{ \Delta  W_\text{rad}^{(3)} }{T} = {\bf E}_\omega^\dagger [ {\boldsymbol \sigma} (\omega, 0) + {\boldsymbol \sigma}^\dagger (\omega, 0) ] {\bf E}_\omega {\bf E}_0^{},
\label{DeltaW-3rd}
\end{align}
Therefore from Eq.~(\ref{DeltaW-3rd}) we see that while a pure monochromatic electric field does not contribute to the power at the third order, there is a non-zero contribution to the work performed by the circuit and the radiation at third order when the DC and the oscillating electric are concomitantly present. The contributions from the second order currents can allow the system to act either as a {\it solar cell}, when the energy is transferred onto the DC circuit, $\Delta W_\text{circ} < 0$, or as a {\it light amplifier} when it is transferred onto the radiation, $\Delta W_\text{rad} < 0$, but such negative work should always be compensated by a positive work to abide by the second law of thermodynamics from Eq.(\ref{DeltaW-non-negative}). We will call a rectification mechanism {\it dissipationless} if the third order contribution to total power as defined in Eq.(\ref{DeltaW-non-negative}) arising from such mechanism vanishes, and we will call it {\it dissipative} if it does not. As we will see, the fact that a rectification mechanism allows for a rectified current within the optical gap of a material is not a sufficient condition for it to be dissipationless, and in fact, we find that generically such in-gap mechanisms are dissipative. We will show that, one specific example of these dissipative mechanisms that allows for in-gap is the semiclassical intra-band Jerk effect in metals~\cite{matsyshyn2019nonlinear}. On the other hand, we will demonstrate that the CPGE associated with the Berry-dipole driven non-linear Hall effect allows for a non-zero current within the transparency region and that it is also a dissipationless mechanism for current rectification in the ideal intra-band limit in which the frequency is much smaller than the optical gap $\Delta_0$.

Let us now specialize our discussion to the effects in metals. To focus on the intra-band effects, we imagine that the inter-band optical gap, $\Delta_0$, is sent to infinity, $\Delta_0 \to \infty$.  At zero temperature and in the ideal limit of vanishing carrier relaxation rates ($\Gamma \to 0$), the metal will have a transparency region in $\omega$ where the dissipative part of the conductivity would vanish as follows:
\begin{align}
{\boldsymbol \sigma} (\omega) + {\boldsymbol \sigma}^\dagger (\omega) = \frac{2 \Gamma}{\omega^2} \pmb{\mathbb D}, 
\quad
\Gamma \ll \omega 
\label{Dissipative-1st}
\end{align}
where $\pmb{\mathbb D}$ is the Drude weight tensor (taken to be symmetrized). Here $\Gamma$ is the relaxation rate that will be defined in a more explicit microscopic form in the Section~\ref{QuantumFormalism} below. From the above we see that in the limit $\Gamma \to 0$ the energy absorption by the material becomes vanishingly small for the frequencies within this optical gap. On the other hand, within this same frequency range, the metal can have finite rectification and as we will see also a non-zero 3rd order contribution to the power as defined from Eq.(\ref{DeltaW-3rd}) that remain finite in the limit of $\Gamma \to 0$. Combining Eqs.~(\ref{DeltaW-2nd}), (\ref{DeltaW-3rd}), and (\ref{Dissipative-1st}), then one would obtain that the leading contributions to the total power are:
\begin{align}
\frac{ \Delta W }{T} = \frac{1}{\Gamma} {\bf E}_0^\text{T} \pmb{\mathbb D} {\bf E}_0^{} + \frac{2\Gamma}{\omega^2} {\bf E}_\omega^\dagger \pmb{\mathbb D} {\bf E}_\omega + {\bf E}_0^\text{T} \pmb{\mathbb K} (\omega) {\bf E}_\omega {\bf E}_\omega^* + \cdots ,
\label{DeltaW-2nd-3rd}
\end{align}
where we replaced ${\boldsymbol \sigma} (0)\to \pmb{\mathbb D}/ \Gamma$, and the sub-leading terms would contain terms of orders, e.g., $O({\bf E}_0^3)$, $O({\bf E}_\omega^4)$, $O({\bf E}_0^2 {\bf E}_\omega^2)$. We have introduced the tensor $\pmb{\mathbb K}(\omega)$ which captures the 3rd order contribution to the total work, and can be obtained from the second order conductivity as follows:
\begin{align}
{\mathbb K}_{a b c} (\omega) = \sigma_{a b c} (\omega, -\omega) + \sigma_{a b c} (0,-\omega) + \sigma_{c a b} (0,\omega) .
\label{K-tensor}
\end{align}
When the tensor $\pmb{\mathbb K}(\omega)$ is non-zero, the rectification process leads to a non-zero contribution to the total work, and, therefore, also to the total heat transfer. As a result, a rectification mechanism operating at a given $\omega$, will be irreversible or dissipative (namely contributing to the entropy change ) if $\pmb{\mathbb K}(\omega) \neq 0$, and it will be reversible or dissipationless (namely not contributing to the entropy change) if $\pmb{\mathbb K}(\omega)=0$.

Notice from Eq.(\ref{DeltaW-2nd-3rd}) that if we had not included the second term accounting for the small but finite residual light absorption arising from Eq.(\ref{Dissipative-1st}), the power in Eq.(\ref{DeltaW-2nd-3rd}) could be made negative for perturbatively small electric fields ${\bf E}_0$, violating the second law of thermodynamics. In fact the minimum of the power as a function ${\bf E}_0$ is obtained for  ${\bf E}_0^\text{min}= - \Gamma \pmb{\mathbb D}^{-1} \pmb{\mathbb K}(\omega) {\bf E}_\omega {\bf E}_\omega^* / 2$ (which is small by virtue of the smallness of $\Gamma$ and ${\bf E}_\omega$), and is given by:
\begin{align}
\frac{ \Delta W }{T}\Big|_\text{min} & = 
\frac{2\Gamma}{\omega^2} {\bf E}_\omega^\dagger \pmb{\mathbb D} {\bf E}_\omega \nonumber\\
& - \frac{\Gamma}{4} [\pmb{\mathbb K} (\omega) {\bf E}_\omega {\bf E}_\omega^*]^\text{T} \pmb{\mathbb D}^{-1} \pmb{\mathbb K} (\omega) {\bf E}_\omega {\bf E}_\omega^* + \cdots .
\label{DeltaW-2nd-3rd-Min}
\end{align}
We see in the above that the second term containing the power arising from dissipative second order processes with nonzero $\pmb{\mathbb K}(\omega)$, is manifestly negative, because the Drude weight is a positive definite tensor. The first term in Eq.(\ref{DeltaW-2nd-3rd-Min}) however is perturbatively larger than the second term and guarantees the positivity of the total power in the perturbative regime. This is the term arising from the small residual light absorption from Eq.(\ref{Dissipative-1st}). Therefore we conclude that the Joule heating term alone is not enough to perturbatively enforce the positivity of the total power, and a small but finite radiation absorption must be present and coexist with the dissipative rectification processes when these induce in-gap photo-currents in order to abide by the second law of thermodynamics, in contrast to the claims in Ref.~\cite{onishi2022photovoltaic}.

\section{Simplified Boltzmann description}
\label{BoltzmannDescription}
To illustrate the above considerations in detail within a simplified model, we consider the single band Boltzmann description within the relaxation-time description employed in Ref.~\cite{sodemann2015quantum}. While this might appear to be a simple-minded treatment~\footnote{We note that this description does not include the correction to the Berry curvature introduced in Ref.~\cite{gao2014field}, which can be neglected in the limit in which the interband energy separation is sent to infinity $\Delta_0 \to \infty$, while keeping the intraband Berry curvature finite so that a projection into a single band is justified (see Ref.~\cite{matsyshyn2019nonlinear}) where one recovers the familiar expression for the Berry phase induced anomalous velocity~\cite{xiao2010berry}.}, in Sections~\ref{QuantumFormalism} and \ref{QuantumResults} we will demonstrate that its predictions are recovered within a fully microscopic description of the system coupled to a heat bath in the intraband limit of $\Gamma \ll \omega \ll \Delta_0$. In the simplified Boltzmann description, the electric current density is: 
\begin{equation}\label{totcurB}
    {\bf j} (t) = \int_{\bf k}  f \left[ \partial_{\bf k} \epsilon + {\bf \Omega} \times {\bf E}(t) \right], 
\end{equation}
where $\int_{\bf k} \equiv \int \text{d}{\bf k} / (2\pi)^d$, $\epsilon$ is the dispersion relation and ${\bf \Omega}$ is the Berry curvature of the band. The electron distribution function $f$ satisfies the Boltzmann equation: 
\begin{equation}
    \partial_t f + {\bf E}(t) \cdot \partial_{\bf k} f = \Gamma(f_0 - f),
\end{equation}
where $\Gamma$ is a relaxation rate and $f_0$ is the equilibrium Fermi-Dirac distribution. Because the anomalous velocity is always orthogonal to the electric field, we immediately see that the work associated with the anomalous current is zero:
\begin{equation}\label{BCwork}
    {\bf j}^\text{anom} (t) \cdot {\bf E}(t) = \int_{\bf k} f [{\bf \Omega} \times {\bf E}(t)] \cdot {\bf E}(t) = 0
\end{equation}

To linear order in electric fields, the electric current is
\begin{equation}
    {\bf j}^{(1)}(\omega) = \frac{1}{\Gamma +i\omega}\pmb{\mathbb D}{\bf E}_\omega+\boldsymbol{\cal F}\times {\bf E}_\omega,
\end{equation}
where $\boldsymbol{\cal F} = \int_{\bf k}   f_0 {\bf \Omega}$ is average of the Berry curvature over the occupied states and $ {\mathbb D}_{a b} = \int_{\bf k} f_0 \partial_a \partial_b \epsilon $ is the Drude weight tensor with $\partial_a \equiv \partial_{k_a}$. It is easy to see that neither the circuit nor the radiation perform work on the electrons via the Berry curvature to leading order of perturbation theory:
\begin{equation}\label{work2}
\Delta W_{\text{BC}}^{(2)} = \Delta W_{\text{circ,BC}}^{(2)} = \Delta W_{\text{rad,BC}}^{(2)}=0.
\end{equation} 

To leading order all the energy transfer to the electrons arises from the Drude weight. More specifically, the work performed by the radiation (radiation energy absorption) and the circuit (Joule heating effect) on the electrons to leading order are given by:
\begin{align}
    \frac{\Delta W_{\text{rad}}^{(2)} }{T}&={\bf j}^{(1)}(\omega)\cdot{\bf E}_{\omega}^*+\text{\text{c.c.}}=\frac{2\Gamma}{\Gamma^2+\omega^2}{\bf E}_\omega\pmb{\mathbb D}{\bf E}^*_\omega,\label{phabs}\\
    \frac{\Delta W_{\text{circ}}^{(2)}}{T} &={\bf j}^{(1)}(0)\cdot{\bf E}_{0}=\frac{1}{\Gamma}{\bf E}_0\pmb{\mathbb D}{\bf E}_0.
\end{align}
The sum of the two terms above gives rise to the second order part of the expression in Eq.~(\ref{DeltaW-2nd-3rd}) in the limit $\Gamma\ll \omega$.

We will now describe the contributions to the next order in perturbation theory. Within the current Boltzmann approach, there are two different mechanism contributing to second order conductivities: the semiclassical Jerk term, arising from the non-parabolicity of the band dispersion, and the non-linear Hall effect, arising from the Berry curvature dipole (BCD)~\cite{sodemann2015quantum,matsyshyn2019nonlinear}. The second order conductivities introduced in Eq.~(\ref{DeltaW-3rd}) can be then separated into Jerk and BCD contributions as follows (for details, see Appendix A):
\begin{align}
    &\sigma_{a b c}^\text{Jerk}(\omega,-\omega) = \frac{2}{\Gamma^2+\omega^2}{\mathbb J}_{a b c},
    \nonumber\\
    &\sigma_{a b c}^\text{BCD}(\omega,-\omega) =\sum_d \left(\frac{\epsilon_{a d c}{\cal D}_{b d}}{\Gamma+i\omega}+\frac{\epsilon_{a d b}{\cal D}_{c d}}{\Gamma-i\omega} \right),
    \nonumber\\
    &\sigma_{a b c}^\text{Jerk}(\omega,0) = \frac{2+i\omega/\Gamma}{(\Gamma+i\omega)^2}{\mathbb J}_{a b c},
    \nonumber\\    
    &\sigma_{a b c}^\text{BCD}(\omega,0) = \sum_d \left( \frac{\epsilon_{adc}{\cal D}_{bd}}{\Gamma+i\omega}+\frac{1}{\Gamma}\epsilon_{adb}{\cal D}_{cd} \right),
    \label{Boltz-sigma}
\end{align}
for $j_c (\omega_1+\omega_2) = \sigma_{cab}(\omega_1,\omega_2)E_{\omega_1, a} E_{\omega_2, b}$, where ${\mathbb J}_{abc} = \int_{\bf k} f_0 \partial_a \partial_b \partial_c \epsilon $ is the Jerk tensor, ${\cal D}_{a b} = \int_{\bf k} f_0 \partial_a \Omega_b $ is the Berry dipole tensor, $\epsilon_{abc}$ is the Levi-Civita symbol, and $E_{\omega, a} \equiv {\bf E}_\omega \cdot {\bf e}_a$. We therefore see that both the Jerk and the Berry dipole indeed give rise to a finite rectification conductivity when the frequency resides within the optical gap, even in the limit of $\Gamma\rightarrow0$. Despite this shared interesting feature, there are, however, some key differences between these mechanisms even at the level of rectification conductivities. One is that the Jerk rectification conductivity tensor, $ \pmb{\mathbb J}$, vanishes in time reversal invariant systems while the Berry dipole tensor remains finite. Moreover, inside the gap and for $\Gamma\rightarrow0$, the real part of the BCD rectification conductivity vanishes, while the real part of the Jerk rectification conductivity remains finite. This implies that in this limit the Jerk mechanism leads to in-gap current rectification driven by linearly polarized light, while the BCD in-gap current rectification requires light with a non-zero degree of circular polarization. Finally, we also see a distinct scaling with frequency, with the BCD and Jerk decaying as $1/\omega$ and $1/\omega^2$ away from the Drude peak, respectively.

After substituting Eqs.(\ref{Boltz-sigma}) into Eq.(\ref{DeltaW-3rd}) one can then obtain the third order contributions to the circuit and radiation powers, which are given by:
\begin{multline}
\frac{\Delta W ^{(3)}_\text{rad}}{T} =\frac{4\Gamma^2}{(\Gamma^2+\omega^2)^2}{\bf E}_0\pmb{\mathbb J} {\bf E}_\omega{\bf E}_\omega^*-\\2 {\bf E}_0\cdot \text{Re}\left[\frac{\boldsymbol{\cal D} {\bf E}_{\omega}\times{\bf E}_\omega^*}{\Gamma+i\omega} \right]  ,
\end{multline}
\begin{multline}
\frac{\Delta W ^{(3)} _\text{circ}}{T} =\frac{1}{\Gamma^2}{\bf E}_0\pmb{\mathbb J} {\bf E}_0{\bf E}_0+\frac{2}{\Gamma^2+\omega^2}{\bf E}_0\pmb{\mathbb J} {\bf E}_\omega{\bf E}_\omega^*+\\2 {\bf E}_0\cdot \text{Re}\left[\frac{\boldsymbol{\cal D} {\bf E}_{\omega}\times{\bf E}_\omega^*}{\Gamma+i\omega} \right].
\end{multline}

Therefore, we see that there are contributions to the individual circuit and radiation works from both the BCD and Jerk terms. Notice, however, that the contributions of the BCD to the circuit and radiation work are exactly opposite to each other and therefore disappear in the net work, as expected from Eq.~(\ref{BCwork}). In contrast the Jerk term contributes to the total work, and therefore the Jerk term is dissipative according to the general considerations of Section~\ref{Thermodynamics}. More specifically the tensor $\pmb{\mathbb K}(\omega)$ introduced in Eq.~(\ref{DeltaW-2nd-3rd}), directly depends on ${\mathbb J}$ (for details, see S.I.A.): 
\begin{equation}\label{PTrestr}
\pmb{\mathbb K}^\text{Boltz}(\omega)=
\frac{6\Gamma^2+2\omega^2}{(\Gamma^2+\omega^2)^2}\pmb{\mathbb J} .
\end{equation}

When ${\mathbb J}\neq0$, $\pmb{\mathbb K}^\text{Boltz}(\omega)$ approaches a non-zero limit for $\Gamma \ll\omega$. Following the general discussion of Section~\ref{Thermodynamics}, we therefore see that a small but finite radiation absorption from the $1/\omega^2$ tail of the Drude peak in Eq.~(\ref{phabs}) necessarily needs to accompany this mechanism in order to not violate the second law of thermodynamics, see Eq.(\ref{DeltaW-2nd-3rd-Min}).

\section{Quantum description with the heat bath --- formalism}
\label{QuantumFormalism}
To investigate to what extent the Boltzmann description we discussed above is valid in capturing microscopic irreversible processes, we construct a fully microscopic quantum description of the crystalline electronic system coupled to a heat bath and subject to a spatially uniform but time dependent vector potential. 

As we will demonstrate, the microscopic description in this section agrees with the simpler Boltzmann description in the limit of $\Gamma \ll \omega \ll \Delta_0$, where $\Delta_0$ is the scale controlling the inter-band optical gap, and thus it serves as a validation of the previous description. But in addition, the full quantum description will allow us to also describe the corrections that appear for frequencies that are comparable to the interband optical gap.

We will use a model of a non-interacting free fermionic bath. This model is in the same class of those non-interacting fermionic models often described within the Keldysh formalism~\cite{nagaosa2017concept,gerchikov1989theory,fregoso2013driven,kamenev2004many,johnsen1999quasienergy,jauho1994time,kohler2005driven,matsyshyn2021rabi}. Here we will provide a description of these baths that avoids the need of second quantization and Keldysh Greens functions (but which is equivalent). 

\begin{figure}
\includegraphics[width=0.49\textwidth]{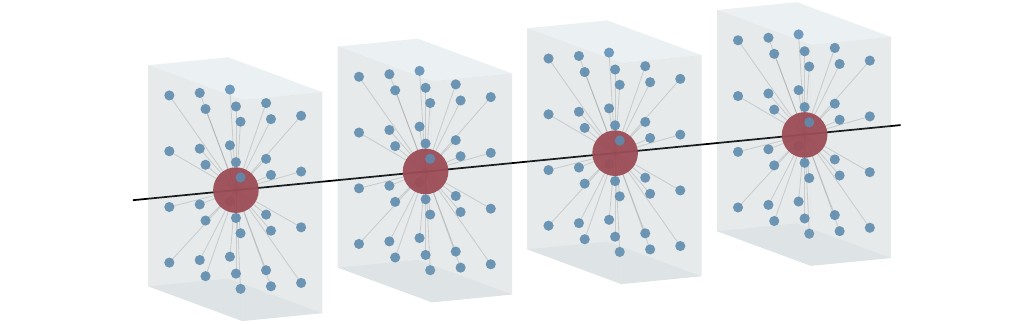}
\caption{Schematic of the crystalline electronic system described by a tight binding model with physical sites (red balls) which are tunnel coupled (solid lines) among themselves, and with their own identical fermionic bath (blue balls).
}
\label{fig-system-bath}
\end{figure}

As depicted in Fig.~\ref{fig-system-bath}, we take a model of the bath in which the system sites are tunnel coupled to a collection of identical bath sites. Thus the system plus bath form a large tight-binding model as a whole. The single particle Hilbert space including the system and the bath can be then decomposed into a direct sum of system and bath subspaces, 
namely their Hamiltonian and states have block form as follows:
\begin{align}
H (t) = \begin{bmatrix}
H_S (t) & H_{SB} \\
H_{SB}^\dagger & H_B
\end{bmatrix} ,
\quad
\psi(t) = \begin{bmatrix}
\psi_S (t) \\
\psi_B (t)
\end{bmatrix} .
\label{Full-Hamiltonian}
\end{align}
The crystalline electronic system is described by a periodic tight-binding Hamiltonian together with the perturbation from the time dependent vector potential 
\begin{align}
H_S(t) & = H_0 + V(t) 
\nonumber\\
& = \sum_n \epsilon_{n}^{} | \chi_n \rangle \langle \chi_n | + \sum_{m n} V_{m n}^{} (t) | \chi_m \rangle \langle \chi_n | , 
\end{align}
where $| \chi_n \rangle$ and $\epsilon_n$ are unperturbed system state and energy with $m,n$ being general indices denoting wave vector, orbital or spin degrees of freedom, while
\begin{align}
V(t) = H_0 ({\bf k} - {\bf A}(t) ) - H_0({\bf k})  . 
\end{align}
The bath reads $H_B = \sum_{n, i} \varepsilon_i^{} \, | \varphi_{n,i} \rangle \langle \varphi_{n,i} |$ with $| \varphi_{n,i} \rangle$ being bath state coupled to the system state $| \chi_n \rangle$ and $\varepsilon_i$ its energy. For simplicity, we set the tunnel coupling $\lambda$ between any system state and bath state to be identical such that $H_{SB} = \lambda \sum_{n, i} | \chi_{n} \rangle \langle \varphi_{n, i} | $. This model is identical to that employed in Refs.~\cite{morimoto2016topological,matsyshyn2021rabi}

From Eq.(\ref{Full-Hamiltonian}), we obtain the coupled Schr\"{o}dinger equations for system and bath states: $i \dot{\psi}_S (t) = H_S (t) \psi_S (t) + H_{SB} \psi_B (t) $ and $i \dot{\psi}_B (t) = H_{SB}^\dagger \psi_S (t) + H_{B} \psi_B (t)$ (we set $\hbar = 1$ throughout the paper). By inserting the second equation into the first one, one can formally eliminate the bath state $\psi_B (t)$ and obtain an integro-differential equation that generalizes the Schr\"{o}dinger equation for the open system $\psi_S (t)$. Its solutions only depend on initial states of the bath and the system, $\psi_B(t_0)$ and $\psi_S(t_0)$. 
Importantly, we now assume the fermionic bath initially is in a thermal state with an equilibrium Fermi-Dirac distribution, namely
\begin{align}
& \rho_B (t_0) = \sum_{n,i} f_0 (\varepsilon_i) | \varphi_{n,i} \rangle \langle \varphi_{n,i}| ,
\nonumber\\
& f_0 (\varepsilon_i) = \frac{1}{\exp[\beta_0 (\varepsilon_i - \mu_0 )]+1},
\end{align}
in which $\mu_0$ is the chemical potential of and $\beta_0 = 1/ k_B T_0 $ denotes the temperature of the bath, respectively, and we send the initial time to minus infinity $t_0 \to -\infty$. It is possible then to obtain the density matrix of the system $\rho_S (t) = \sum_{n=0}^\infty \rho^{(n)} (t) $ perturbatively in terms of $V(t)$.

The bath is taken into a thermodynamic limit in which its spectrum of energies $\varepsilon_i$ becomes continuum and is described by a density of states:
\begin{align}
\nu_B(\omega) = \sum_i \delta(\omega-\varepsilon_i)  .
\end{align}
For simplicity we take an ideal bath with a flat and infinitely broad spectrum, namely, we take its density of states to be a constant, $\nu_B(\omega) = \nu_0$. The relaxation rate scale associated with the bath will then be:
\begin{align}
\Gamma = \nu_0 \lambda^2/2 .
\end{align}
With the above simplifications it is possible to find relatively simple closed expressions for the density matrix of the system expanded in powers of the time dependent perturbation. We obtain the density matrix expansions to the zeroth order in $V(t)$:
\begin{align}
\rho_{m n}^{(0)} = \delta_{m n}
\int_{\omega_b}
\frac{ 2 \Gamma }
{\omega_b^2 + \Gamma^2}
f_0(\epsilon_{m} + \omega_b) ,
\label{rho_0}
\end{align}
where we used shorthand notations $\int_{\omega} \equiv \int_{-\infty}^{\infty} \text{d} \omega / 2 \pi$ as well as $\epsilon_{n m} \equiv \epsilon_{n} - \epsilon_{m}$. Here the subscripts $m$, $n$ are generic and include both momentum and band (e.g., orbital, spin, valley) indices. The above distribution accounts for the broadening of the energy levels of the system due to its coupling to the bath, and reduces to the ideal Fermi-Dirac distribution in the limit of $\Gamma \to 0$. Additionally, expanding $V(t)$ to the first order and the second order, we obtain
\begin{align}
\begin{split}
\rho_{m n}^{(1)} (t) =
\int_{\omega} e^{-i \omega t}
V_{m n} (\omega) \, \tilde{\rho}_{m n}^{(1)} (\omega) ,
\end{split}
\nonumber\\
\begin{split}
\rho_{m n}^{(2)} (t) =  \int_{\omega_1} \int_{\omega_2} e^{-i (\omega_1 + \omega_2) t} \sum_{l} V_{m l} (\omega_1) V_{l n} (\omega_2)
\\
\times \tilde{\rho}_{m l n}^{(2)} (\omega_1, \omega_2),
\end{split}
\end{align}
in which we have
\begin{align}
& \tilde{\rho}_{m n}^{(1)} (\omega) = \int_{\omega_b} \frac{ 2\Gamma } {\omega_b^2 + \Gamma^2} \
\frac{ f_0(\epsilon_{n} + \omega_b ) - f_0(\epsilon_{m} - \omega_b )}
{ \omega + \omega_b + \epsilon_{n m}  + i \Gamma },
\nonumber \\
& \tilde{\rho}_{m l n}^{(2)} (\omega_1,\omega_2) = \int_{\omega_b} \frac{ 2 \Gamma }
{\omega_b^2 + \Gamma^2} \times
\nonumber \\
& \bigg[ \frac{f_0(\epsilon_{m} - \omega_b )}{ (\omega_1 + \omega_2 + \omega_b + \epsilon_{n m} + i \Gamma) (\omega_1 + \omega_b +  \epsilon_{l m} + i \Gamma )}
\nonumber \\
& + \frac{f_0(\epsilon_{n} + \omega_b )}{ ( \omega_1 + \omega_2 + \omega_b + \epsilon_{n m} + i \Gamma ) (\omega_2 + \omega_b + \epsilon_{n l} + i \Gamma )}
\nonumber \\
& - \frac{ f_0(\epsilon_{l} + \omega_b ) }{ ( \omega_1 + \omega_b +  \epsilon_{l m} + i \Gamma ) ( \omega_2 - \omega_b +  \epsilon_{n l} + i \Gamma ) }
\bigg] .
\end{align}

In the case of our interest, however, the perturbation $V(t)=H_0({\bf k}-{\bf A})-H_0 ({\bf k})$ itself has a non-linear dependence on ${\bf A}(t)$. In order to calculate the current density along with the work performed by the circuit and radiation, we also need expansions of the perturbation $V(t)$ in terms of ${\bf A}$ as
\begin{align}
V^{(n)} (t) = \sum_{a_1 \cdots a_n} & \frac{(-1)^n}{n!} \frac{ \partial^n H_0 ({\bf k})}{\partial_{a_1} \cdots \partial_{a_n} }
\nonumber\\
& \times A_{\omega_1, a_1} \cdots A_{\omega_n, a_n} ,
\end{align}
in which $a_n = x,y,z$ stands for spatial indices, $\partial_a \equiv \partial_{k_a}$, and $A_{\omega,a} = {\bf A}_{\omega} \cdot {\bf e}_a$~\footnote{Expanding perturbation (and current operator below) by derivatives requires applying unitary transformations $\exp[\mp i {\bf k} \cdot ({\bf x}_\alpha - {\bf x}_\beta)]$ before and after derivatives, with ${\bf x}_{\alpha}$ the position for the $\alpha$-th atom in the unit cell. Here we assumed all atoms have the same position for simplicity~\cite{simon2020contrasting}}.
Its Fourier transforms are given by:
\begin{align}
V^{(n)} (\omega) =\sum_{a_1 \cdots a_n}
\delta(\omega - {\textstyle \sum}_n \omega_n) & A_{\omega_1, a_1} \cdots A_{\omega_n, a_n}
\nonumber \\
& \times \tilde{V}^{(a_1\cdots a_n)},
\end{align}
where we used the notation
\begin{align}
\tilde{V}^{(a_1\cdots a_n)}
\equiv 2\pi \frac{(-1)^n}{n!} \frac{ \partial^n H_0 ({\bf k})}{\partial_{a_1} \cdots \partial_{a_n} } .
\end{align}

Similarly, for the current operator $J_a (t) = - \partial H_0 ({\bf k}-{\bf A} )/\partial A_a = \partial H_0 ({\bf k}-{\bf A} )/\partial k_a $, its expansions in ${\bf A}$ are
\begin{align}
J_a^{(0)} = \partial_a H_0 ({\bf k}) ,
\quad
J_a^{(n)} (t) = \partial_a V^{(n)} (t) .
\end{align}
Their Fourier transforms are $J_a^{(0)} (\omega) = 2\pi \partial_a H_0 ({\bf k})$ and
\begin{align}
J_a^{(n)} (\omega) = \sum_{a_1 \cdots a_n} \delta(\omega - {\textstyle \sum}_n \omega_n) & A_{\omega_1, a_1} \cdots A_{\omega_n, a_n}
\nonumber \\
& \times \tilde{J}_a^{(a_1 \cdots a_n)},
\end{align}
in which we defined
\begin{align}
\tilde{J}_a^{(a_1 \cdots a_n)} \equiv  \partial_a \tilde{V}^{(a_1\cdots a_n)} .
\label{J_n}
\end{align}

From the Eqs.~(\ref{rho_0}) to (\ref{J_n}), we are able to calculate current densities and the corresponding conductivity tensors to linear order in ${\bf A}$,
\begin{align}
\begin{split}
j_a^{(1)} (\omega) = \sigma_{a b}(\omega) E_{\omega, b} ,
\end{split}
\nonumber\\
\begin{split}
\sigma_{ab} (\omega) = \frac{i}{ \omega}
\sum_{mn} \Big[ \tilde{V}_{mn}^{(b)} \tilde{\rho}_{mn}^{(1)} (\omega) J_{a,nm}^{(0)} + \rho_{mn}^{(0)} \tilde{J}_{a, nm}^{(b)} \Big] ,
\end{split}
\label{sigma_ab}
\end{align}
and those to the second order,
\begin{align}
& j_a^{(2)} (\omega_1 + \omega_2) = \sigma_{a b c}(\omega_1, \omega_2) E_{\omega_1, b} E_{\omega_2, c} , 
\nonumber\\
& \sigma_{abc} (\omega_1, \omega_2) = \frac{i}{\omega_1}
\frac{i}{\omega_2} \sum_{mln}
\Big[ 
\tilde{V}_{ml}^{(b)} \tilde{V}_{ln}^{(c)} \tilde{\rho}_{mln}^{(2)} (\omega_1,\omega_2) J_{a,nm}^{(0)}
\nonumber\\
& + \tilde{V}_{mn}^{(bc)} \tilde{\rho}_{mn}^{(1)} (\omega_1+\omega_2) J_{a,nm}^{(0)} +
\tilde{V}_{mn}^{(b)} \tilde{\rho}_{mn}^{(1)} (\omega_1) \tilde{J}_{a, nm}^{(c)}
\nonumber\\
& + \rho_{mn}^{(0)} \tilde{J}_{a, nm}^{(bc)} \Big] +
\Big( \
\begin{matrix}
b \leftrightarrow c
\\
\omega_1 \leftrightarrow \omega_2
\end{matrix} \
\Big).
\label{sigma_abc}
\end{align}
\begin{figure*}
\includegraphics[width=0.99\textwidth]{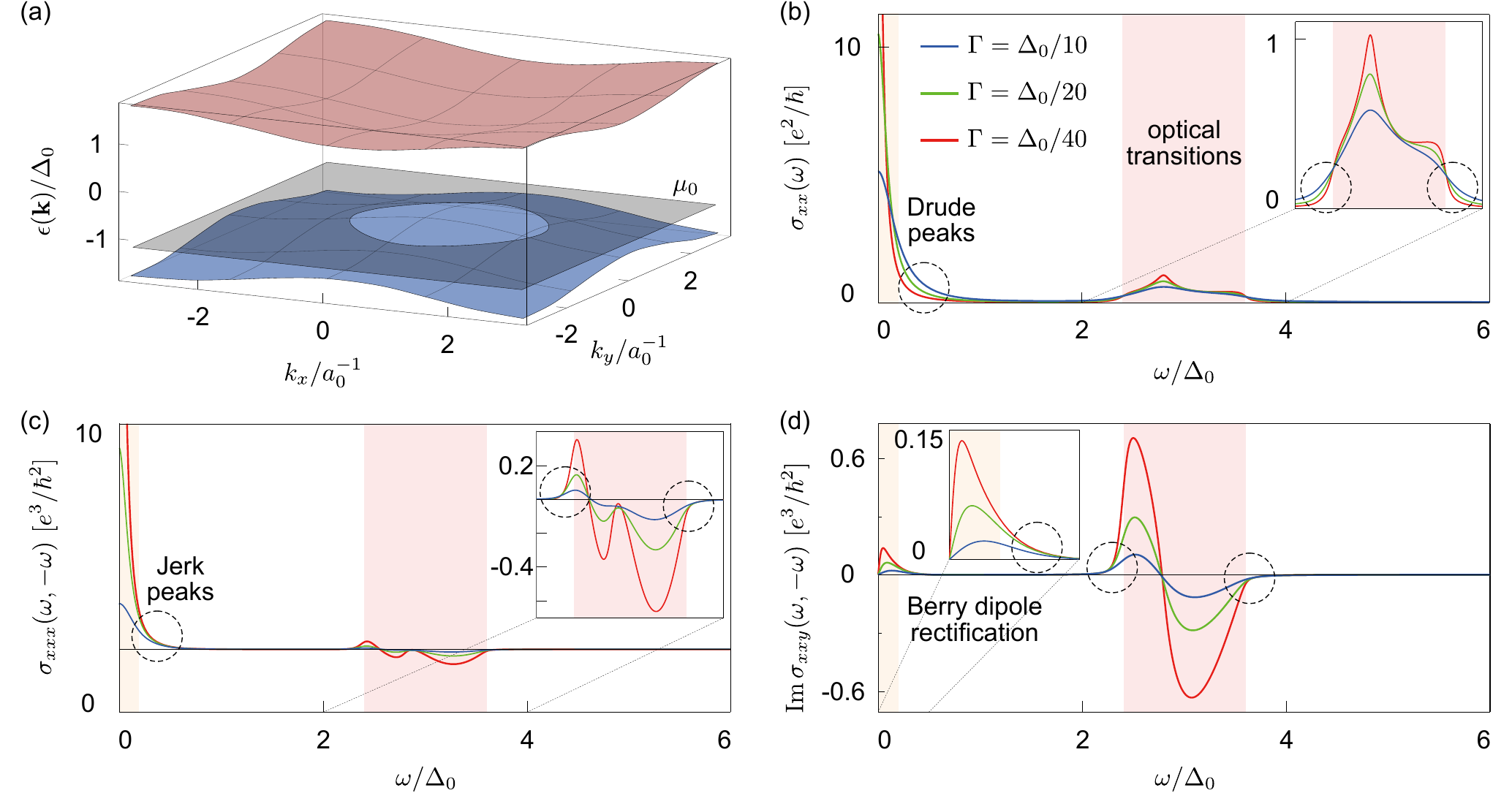}
\caption{(a) Dispersion relations for conduction band (red) and valence band (blue) of the time reversal breaking Hamiltonian $h({\bf k})$ in its Brillouin zone, for a chemical potential $\mu_0$ (gray plane) crossing the valence band. (b), (c), and (d) Conductivities $\sigma_{xx} (\omega)$, $\sigma_{xxx} (\omega,-\omega)$, and $\text{Im} \sigma_{xxy}(\omega, -\omega)$ for $\Gamma = \Delta_0 / 10$ (blue lines), $\Gamma = \Delta_0 / 20$ (green lines), and $\Gamma = \Delta_0 / 40$ (red lines). Light red areas denote the energy range in which optical transitions between the conduction and valence bands are allowed, while all other areas are gap regions. Light orange areas denote the energy range between the chemical potential and top of the valence band. The insets are zoomed-in views for regions of small values. Dashed circles highlight trends of in-gap conductivities as $\Gamma$ decreases.  Parameters used: $\Delta_0 = 1$, $a_0 = 1$, $t_x = \Delta_0 / 5$, $t_y = \Delta_0 / 6$, $\phi_{x,1}= \pi/5$,  $\phi_{y,1}= \pi/7$, $m=\Delta_0 / 5$, $\phi_{x,2}=\pi/13$, $\phi_{y,2}=\pi/11$; $\mu_0 = -6 \Delta_0/5$, $\beta_0 = 100 / \Delta_0$.
}
\label{fig-dispersion-conductivity}
\end{figure*}
We note that one needs to take limits, e.g., $\sigma_{ab} (0) = \lim_{\omega \to 0} \sigma_{ab} (\omega)$ and $\sigma_{abc} (\omega, -\omega) = \lim_{\omega_1 \to -\omega} \sigma_{abc} (\omega, \omega_1) $ when encountering zero frequencies. Using Eq.~(\ref{sigma_abc}), we can then compute arbitrary components of $\pmb{\mathbb K} (\omega)$ from its definition Eq.~(\ref{K-tensor}).

\section{Quantum description with the heat bath --- numerical results}
\label{QuantumResults}
After describing the quantum microscopic formalism, we now consider a specific two-dimensional tight binding model
\begin{align}
h ({\bf k}) = d_x ({\bf k}) \sigma_x + d_y ({\bf k}) \sigma_y + d_z ({\bf k}) \sigma_z ,
\end{align}
in which $\sigma_{x,y,z}$ are Pauli matrices; and we chose the following expressions for the vector parameterizing the Bloch Hamiltonian: $ d_x ({\bf k}) = t_x \sin (k_x a_0 - \phi_{x,1})$ and $d_y ({\bf k}) = t_y \sin (k_y a_0 - \phi_{y,1})$ represent complex nearest neighbour hoppings in $x$- and $y$- directions, and $d_z ({\bf k}) = \Delta_0 + m [ 2 - \cos (k_x a_0 - \phi_{x,2}) - \cos (k_y a_0 - \phi_{y,2}) ]$ with $\Delta_0$ characterizing the band gap size. We set the lattice constant $a_0 = 1$, and choose a generic set of phase factors $\phi_{x(y),1(2)} \neq 0$ such that  $h ({\bf k})$ does not have the time reversal symmetry or any crystalline symmetries. 

Fig.~\ref{fig-dispersion-conductivity}(a) shows a typical dispersion relations for $h({\bf k})$. The gap size between the valence band and the conduction band is approximately $2\Delta_0$ for the parameters chosen for this figure. To focus on effects in metals, we chose a chemical potential $\mu_0$ that crosses the valence band. Using Eqs.~(\ref{sigma_ab}) and (\ref{sigma_abc}), and for different $\Gamma$, we calculated linear and nonlinear conductivities for $h({\bf k})$. Illustrative results are shown in Fig.~\ref{fig-dispersion-conductivity}(b), (c), and (d).

Fig.~\ref{fig-dispersion-conductivity}(b) for the linear conductivity $\sigma_{xx} (\omega)$ shows that our bath has well-behaved current relaxation: there is a non-zero DC conductivity for $\omega \to 0$ and finite $\Gamma$, and at low frequencies, one can observe a Drude peak that becomes sharper as $\Gamma$ decreases. Also, from the inset of Fig.~\ref{fig-dispersion-conductivity}(b), one can see that when the frequency lies in the gap region (white areas), namely regions outside the energy range in which optical transitions are allowed (light red areas), the linear conductivity $\sigma_{xx}(\omega)$ approaches zero as $\Gamma \to 0$ (e.g., see dashed circles).

On the other hand, Fig.~\ref{fig-dispersion-conductivity}(c) and (d) illustrate that there are indeed in-gap rectifications, exemplified by non-vanishing $\sigma_{xxx} (\omega,-\omega)$ and $\text{Im}\,\sigma_{xxx} (\omega,-\omega)$ in the gap region in the limit of $\Gamma \to 0$ (e.g., see dashed circles). This shows that non-vanishing rectification currents are not artefacts of the simpler Boltzmann description.

\begin{figure}
\includegraphics[width=0.49\textwidth]{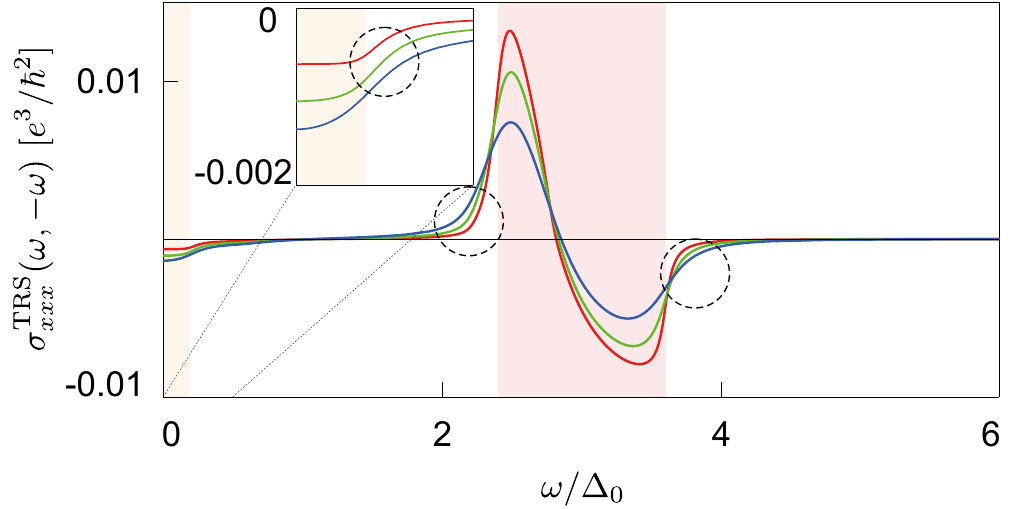}
\caption{Time reversal symmetric rectification conductivity $\sigma_{xxx}^\text{TRS} (\omega,-\omega)$ for $h^\text{TRS} ({\bf k})$ at $\Gamma = \Delta_0 / 10$ (blue lines), $\Gamma = \Delta_0 / 20$ (green lines), and $\Gamma = \Delta_0 / 40$ (red lines). The colored areas, and markers have the same meaning as in Fig.~\ref{fig-dispersion-conductivity}. Parameters used are also the same with those in Fig.~\ref{fig-dispersion-conductivity}. This figure illustrates that this conductivity vanishes within the optical gap in the limit $\Gamma \to 0$.
}
\label{fig-conductivity-TRS}
\end{figure}

To be able to isolate more precisely the in-gap rectification mechanisms that are present in time reversal symmetric systems, we consider the following related model of time reversal symmetric Hamiltonian:
\begin{align}
h^\text{TRS} ({\bf k}) = \begin{bmatrix}
h ({\bf k}) & 0 \\
0 & h^*(-{\bf k})
\end{bmatrix} .
\label{hTRS}
\end{align}
Namely this model is simply made by adding an additional time reversed copy to the earlier time reversal breaking model, making the new model time reversal invariant as a whole. The idea is that this new model is expected to display Berry dipole rectification but no Jerk effect. In fact, for the non-linear conductivity $\text{Im}\,\sigma_{xxy}^\text{TRS} (\omega,-\omega)$, which contains information about the rectification of circularly polarized light from the Berry dipole, one can verify that one simply obtains twice the previous result, namely that:
\begin{align}
\text{Im}\,\sigma_{xxy}^\text{TRS} (\omega,-\omega) = 2 \, \text{Im}\,\sigma_{xxy} (\omega,-\omega) .
\end{align}
However, except for this special case,  other time reversal symmetric rectification conductivities behave differently from their time reversal breaking counterparts, which we exemplified by plotting  $\sigma_{xxx}^\text{TRS} (\omega,-\omega)$ in Fig.~\ref{fig-conductivity-TRS}. We can see in Fig.~\ref{fig-conductivity-TRS} that, within the gap region, time reversal symmetric conductivity $\sigma_{xxx}^\text{TRS} (\omega,-\omega) \to 0$, as $\Gamma \to 0$ (e.g., see dashed circles). This is in contrast to the nonvanishing of time reversal breaking conductivity $\sigma_{xxx} (\omega,-\omega) $ as $\Gamma$ decreases shown in Fig.~\ref{fig-dispersion-conductivity}(c). Importantly, when $\omega$ is within the gap region, we verified that
\begin{align}
& \lim_{\Gamma \to 0} \sigma_{xxx}^\text{TRS} (\omega, -\omega) = \lim_{\Gamma \to 0} \sigma_{xyy}^\text{TRS} (\omega, -\omega)
\nonumber\\
= & \lim_{\Gamma \to 0} \text{Re}\,\sigma_{xxy}^\text{TRS} (\omega, -\omega) = 0 .
\quad
(\omega \in \text{gap})
\label{no-rectifications}
\end{align}
The above confirms that the rectification conductivities arising from the Jerk mechanism vanish in time reversal symmetric crystals, which is consistent with the conclusion in Eq.~(\ref{Boltz-sigma}). However, we also verified that
\begin{align}
\lim_{\Gamma \to 0} \text{Im}\,\sigma_{xxy}^\text{TRS} (\omega, -\omega) \neq 0 ,
\quad
(\omega \in \text{gap})
\end{align}
namely the rectification conductivity form the Berry dipole does not vanish for both time reversal breaking and time reversal symmetric crystals [see Eq.~(\ref{Boltz-sigma})].

After validating the existence of in-gap rectifications with and without the time reversal symmetry, we now turn to analyze the third order total power $\Delta W^{(3)} / T$ in the clean limit $\omega \gg \Gamma \to 0$, which is controlled by the $\pmb{\mathbb K} (\omega)$ tensor defined in Eqs.~(\ref{DeltaW-2nd-3rd}) and (\ref{K-tensor}). For the Hamiltonian $h ({\bf k})$ without any symmetries, all components of $\pmb{\mathbb K} (\omega)$ can be nonzero. For simplicity, we assume that ${\bf E}_0 = E_0 {\bf e}_x$ such that we can focus on the components ${\mathbb K}_{xab}$ ($a,b \in x, y$). We note that the realness of the power mandates that ${\mathbb K}_{xxx}, {\mathbb K}_{xyy} \in {\mathbb R}$ are real; and due to symmetrized conductivity tensors [see Eq.~(\ref{symmetrized-sigma})], using Eq.~(\ref{K-tensor}) we have ${\mathbb K}_{xyx} (\omega) = {\mathbb K}_{xxy} (-\omega) = {\mathbb K}_{xxy}^* (\omega)$. Therefore the only four independent components for ${\mathbb K}_{xab}$ ($a,b \in x, y$) are  ${\mathbb K}_{xxx}(\omega)$, ${\mathbb K}_{xyy}(\omega)$, $\text{Re}\,{\mathbb K}_{xxy}(\omega)$ and $\text{Im}\,{\mathbb K}_{xxy}(\omega)$. We note that ${\mathbb K}_{xxx} (\omega)$, ${\mathbb K}_{xyy} (\omega)$ and $\text{Re}\,{\mathbb K}_{xxy} (\omega)$ correspond to the work $\Delta W^{(3)} / T$ performed by linearly polarized radiations; on the other hand, $\text{Im}\, {\mathbb K}_{xxy} (\omega)$ appears exclusively for work related to circularly polarized radiation.
\begin{figure}
\includegraphics[width=0.49\textwidth]{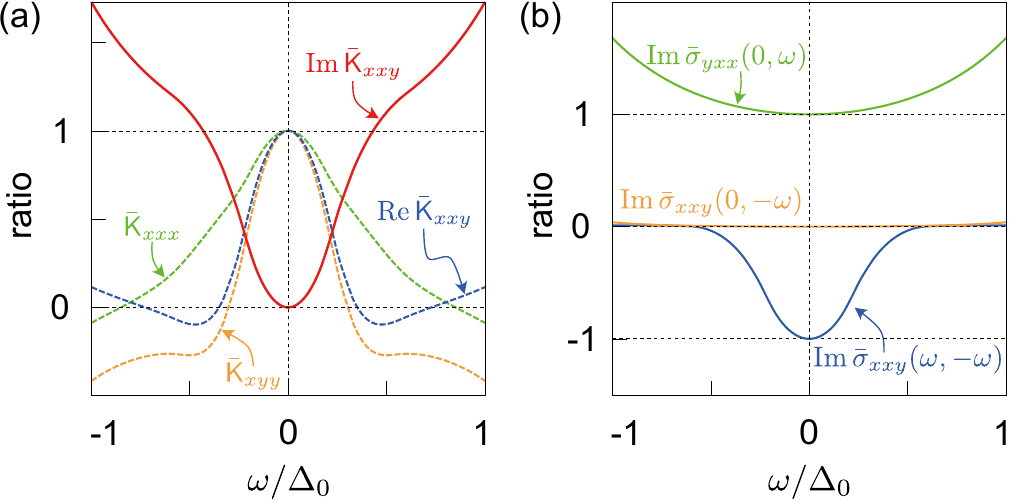}
\caption{(a) Ratios between the microscopic bath description and Boltzmann description of the components of $\lim_{\Gamma \to 0} \pmb{\mathbb K} (\omega)$ which determine the 3rd order contributions to power $\Delta W^{(3)}$ for ${\bf E}_0 = E_0 {\bf e}_x$ in a 2D crystal without the time reversal symmetry or any crystalline symmetries. The fact that these curves approach $1$ when $\omega \ll \Delta_0$, validates the Boltzmann theory in the intra-band limit. When the time reversal symmetry is present so that the Berry dipole is the only intra-band rectification mechanism that is present, we see that only $\text{Im}\,\bar{\mathbb K}_{xxy}$ from the Berry dipole ${\cal D}_{xz}$ persists and remains finite (solid line), while the other three components vanish (dashed lines). The vanishing of all the components of the work tensor $\pmb{\mathbb K}(\omega)$ in the limit $\omega \ll \Delta_0$, indicates that the intra-band Berry dipole rectification is dissipationless. (b) Plots for three conductivities that constitute $\lim_{\Gamma \to 0} \text{Im}\,{\mathbb K}_{xxy} (\omega)$ and are normalized by ${\cal D}_{xz} / \omega$. Parameters used are the same with those in Fig.~\ref{fig-dispersion-conductivity}, while all results are obtained at $\omega \gg \Gamma \to 0$ limit. 
}
\label{fig-tensor}
\end{figure}

To compare in detail the four independent $\pmb{\mathbb K} (\omega)$ components for $h ({\bf k})$ from the quantum bath description and those from Boltzmann formalism, in Fig.~\ref{fig-tensor}(a), we computed the ratios between these components calculated from Eqs.~(\ref{K-tensor}) and (\ref{sigma_abc}), and those from Eq.~(\ref{PTrestr}) in the $\omega \gg \Gamma \to 0$ limit: 
\begin{align}
& \bar{\mathbb K}_{xxx} = \lim_{\Gamma \to 0} 
\frac{ {\mathbb K}_{xxx} (\omega) }{{\mathbb K}_{xxx}^\text{Boltz} (\omega)} ,
\quad
\text{Re}\,\bar{\mathbb K}_{xxy} = \lim_{\Gamma \to 0} \frac{ \text{Re}\,{\mathbb K}_{xxy} (\omega) }{ {\mathbb K}_{xxy}^\text{Boltz} (\omega) },
\nonumber\\
& \bar{\mathbb K}_{xyy} = \lim_{\Gamma \to 0}
\frac{ {\mathbb K}_{xyy} (\omega) }{ {\mathbb K}_{xyy}^\text{Boltz} (\omega) },
\quad
\text{Im}\,\bar{\mathbb K}_{xxy} = \lim_{\Gamma \to 0} \frac{ \text{Im}\, {\mathbb K}_{xxy} (\omega) }{ {\cal D}_{xz} / \omega } ,
\end{align}
where we used [see Eq.~(\ref{Boltz-sigma})]
\begin{align}
- & \lim_{\Gamma \to 0} \text{Im}\, \sigma_{xxy}^\text{Boltz} (\omega,-\omega) = \lim_{\Gamma \to 0} \text{Im}\, \sigma_{yxx}^\text{Boltz} (0,\omega)
\nonumber\\
& = \frac{ {\cal D}_{xz} }{\omega} = \frac{1}{\omega} \int_{\bf k} f_0 \partial_x \Omega_z,
\end{align}
to normalize $\text{Im}\,{\mathbb K}_{xxy} (\omega)$, because within the Boltzmann formalism $\lim_{\Gamma \to 0} \text{Im}\,{\mathbb K}_{xxy}^\text{Boltz} (\omega) = 0$.
These ratios are plotted in Fig.~\ref{fig-tensor}(a) which demonstrate that our results from the quantum theory validate the Boltzmann analysis in the $\omega \ll \Delta_0 \to \infty$ limit, which reproduce exactly the predictions of Boltzmann formalism in this regime.

The microscopic multiband formalism allows us also to characterize the deviations beyond the Boltzmann intraband description when the frequency is comparable with the interband optical gap: for a finite $\omega/\Delta_0$, $\bar{\mathbb K}_{xxx}$, $\bar{\mathbb K}_{xxx}$, and $\text{Re}\, \bar{\mathbb K}_{xxy}$ deviate from the unity; while $\text{Im}\, \bar{\mathbb K}_{xxy}$ becomes nonzero.
Therefore, in general, all four components lead to nonzero $\Delta W^{(3)}/T$ and are dissipative when interband effects are taken into account.

For the time reversal symmetric model from Eq.~(\ref{hTRS}), we performed the same analysis for the four independent $\pmb{\mathbb K} (\omega)$ components. In this case, we found
\begin{align}
\lim_{\Gamma \to 0} 
{\mathbb K}_{xxx}^\text{TRS} (\omega) = \lim_{\Gamma \to 0} 
{\mathbb K}_{xyy}^\text{TRS} (\omega) =
\lim_{\Gamma \to 0} 
\text{Re}\, {\mathbb K}_{xxy}^\text{TRS} (\omega) = 0 ,
\end{align}
namely, in time reversal symmetric crystals and in the $\Gamma \ll \omega < \Delta_0$ limit, total work $\Delta W^{(3)} / T$ related to linearly polarized radiations are zero. Moreover, from Eq.~({\ref{no-rectifications}}), one can conclude that $\Delta W_\text{circ}^{(3)} / T = \Delta W_\text{rad}^{(3)} / T = 0 $ vanish simultaneously in this circumstances.

On the other hand, $\text{Im}\, {\mathbb K}_{xxy}^\text{TRS} (\omega)$ takes the same value as in the time reversal broken model, namely we have that:
\begin{align}
\lim_{\Gamma \to 0} \text{Im}\, \bar{\mathbb K}_{xxy}^\text{TRS} (\omega)
& = \frac{\lim_{\Gamma \to 0} \text{Im}\, {\mathbb K}_{xxy}^\text{TRS} (\omega)}{
{{\cal D}_{xz}^\text{TRS}}}
\nonumber\\
& = \frac{2 \lim_{\Gamma \to 0}  \text{Im}\, {\mathbb K}_{xxy} (\omega)}{
{2 {\cal D}_{xz}}}
\nonumber\\
& = \lim_{\Gamma \to 0}  \text{Im}\, \bar{\mathbb K}_{xxy} (\omega) ,
\end{align}
The above component is illustrated in Fig.~\ref{fig-tensor}(a) using the solid line. From Fig.~\ref{fig-tensor}(a), one can also observe that in the $\Gamma \ll \omega \ll \Delta_0$ limit, $\text{Im}\, {\mathbb K}_{xxy} (\omega) = 0$. Therefore, this demonstrates that indeed for time reversal symmetric crystals the total work performed by circularly polarized radiation via the Berry dipole mechanism exactly vanishes, namely, that this mechanism of rectification is dissipationless in this limit.
The striking point is that this occurs while in-gap rectification conductivity itself remains finite, namely, in this limit $\text{Im}\, \sigma_{xxy}^\text{TRS} (\omega,-\omega) = 2\, \text{Im}\, \sigma_{xxy} (\omega,-\omega) = -2 D_{xz} / \omega $ remains finite as we illustrate in Fig.~\ref{fig-tensor}(b). This means that $\Delta W_\text{circ}^{(3)}/T =  - \Delta W_\text{rad}^{(3)}/T \neq 0$, or, in other words, that it is possible to perform dissipationless energy transfer between the circuit and circularly polarized radiation, in agreement with the considerations of Section~\ref{BoltzmannDescription}.

\section{Applications}
\label{Applications}
In this section we will discuss how the dissipationless nature of the intra-band non-linear Hall effect has a potential to develop highly efficient photovoltaic and light amplification devices. This is because the non-linear Hall effect arising from the Berry curvature dipole, behaves as a ``photovoltaic Demon'', namely it transfers completely the energy from the radiation onto the circuit in a reversible fashion without any energy dissipated onto the heat bath. The BCD effect will necessarily coexist with other dissipative effects, such as Joule heating, and as a result there will always be a net imperfect conversion of energy from the radiation onto the circuit. We will show, however, that the ultimate bound of the efficiency of energy conversion is 100\%, and can be approached when the non-linear Hall effect dominates over the Joule heating and the dissipative photon absorption processes. 

The dissipationless nature of the non-linear Hall effect arises from the fact that the anomalous velocity is orthogonal to the total electric field [see Eq.~(\ref{totcurB})], leading to a perfect cancellation of the radiation and circuit BCD contributions to the total work:
\begin{equation}\label{BDtotwork}
    \Delta W^{(3)}_{\text{BD}}=\Delta W^{(3)}_{\text{circ,BD}}+\Delta W^{(3)}_{\text{rad,BD}}=0.
\end{equation}    
The above is the mathematical statement that the BCD does not produce heat and behaves as a photovoltaic demon that transfers completely the energy between the circuit and the radiation.

The electronic system operates as a solar cell when $\Delta W_\text{rad} > 0$ and $\Delta W_\text{circ} < 0$. In this regime the system absorbs energy from the radiation and transfer it onto the circuit. In the opposite case, it behaves as an amplifier of light, when the energy of the circuit is delivered onto the radiation $\Delta W_\text{rad}< 0$. We therefore introduce two kinds of energy efficiency functions for the two modes of operation of the electronic system:
\begin{gather}
    \eta_\text{Solar} = -\frac{\Delta W_{\text{circ}}}{\Delta W_{\text{rad}}},\text{ for }\Delta W_\text{rad} > 0,\Delta W_\text{circ} < 0, \label{effSolar}\\ \eta_\text{Amp} = -\frac{\Delta W_{\text{rad}}}{\Delta W_{\text{circ}}},\text{ for }\Delta W_\text{rad}< 0,\Delta W_\text{circ} >0.\label{effAmp}
\end{gather}
In the above equations  $\Delta W_\text{circ}$ and $\Delta W_\text{rad}$ are understood to be the respective works of circuit and radiation including all processes, both dissipative and dissipationless. Notice that the second law of thermodynamics from Eq.(~\ref{DeltaW-non-negative}) implies that each of the above efficiencies is always bounded by 1: $\eta\leq 1$. Since the non-linear Hall effect is allowed in time reversal invariant systems~\cite{sodemann2015quantum} and in order to eliminate the dissipative jerk term, from here on we assume that our system has time reversal symmetry leading to ${\mathbb J}=0$. In this case the work to leading 3rd order in electric field is:
\begin{align}
    \frac{\Delta W_{\text{circ}}}{T}&=\frac{1}{\Gamma} {\bf E}_0\pmb{\mathbb D}{\bf E}_0-2{\bf E}_0\cdot \text{Re}\left[\frac{{\bf E}^*_{\omega}\times\boldsymbol{\cal D} {\bf E}_{\omega}}{\Gamma+i\omega} \right],\\
    \frac{\Delta W_{\text{rad}}}{T}&={2\Gamma}\frac{{\bf E}^*_{\omega}\pmb{\mathbb D}{\bf E}_\omega}{\Gamma^2+\omega^2}+2{\bf E}_0\cdot \text{Re}\left[\frac{{\bf E}^*_{\omega}\times\boldsymbol{\cal D} {\bf E}_{\omega}}{\Gamma+i\omega} \right].\label{TRSworkRad}
\end{align}

The first terms of $\Delta W_\text{circ}$ and $\Delta W_\text{rad}$ are the Joule heating effect and the photon absorption from the Drude peak, respectively, which are both dissipative processes. The second terms are the BCD contributions, which we see are exactly opposite, as expected from Eq.(\ref{BDtotwork}). We see also that, for $\omega\gg\Gamma$, the sign of the product of ${\bf E}_0$ with the vector $\mathrm{Im}\left[ {\bf E}^*_{\omega}\times\boldsymbol{\cal D} {\bf E}_{\omega}\right]$, is what determines whether the work can be negative, and therefore the sign of this is what ultimately determines if the system operates as a solar cell or as an amplifier (see the sign of the work done by a radiation from Eq.(\ref{TRSworkRad})). ${\bf E}_0$ is determined by the external circuit, whereas $\mathrm{Im}\left[ {\bf E}^*_{\omega}\times\boldsymbol{\cal D} {\bf E}_{\omega}\right]$ is determined by the radiation. Moreover, $\mathrm{Im}\left[ {\bf E}^*_{\omega}\times\boldsymbol{\cal D} {\bf E}_{\omega}\right]$ is only non-zero when the radiation has a finite degree of circular polarization and reverses direction when the handedness of the polarization is reversed. The intuition behind this product, is that $\mathrm{Im}\left[ {\bf E}^*_{\omega}\times\boldsymbol{\cal D} {\bf E}_{\omega}\right]$ is the direction in which the rectified current would flow when only ${\bf E}_{\omega}$ is present, and therefore we have a solar cell when ${\bf E}_0$ is trying to oppose such current flow and a radiation amplifier when ${\bf E}_0$ is aiding it (which requires the circuit to deliver the energy to sustain this).
\begin{figure}
\includegraphics[width=0.49\textwidth]{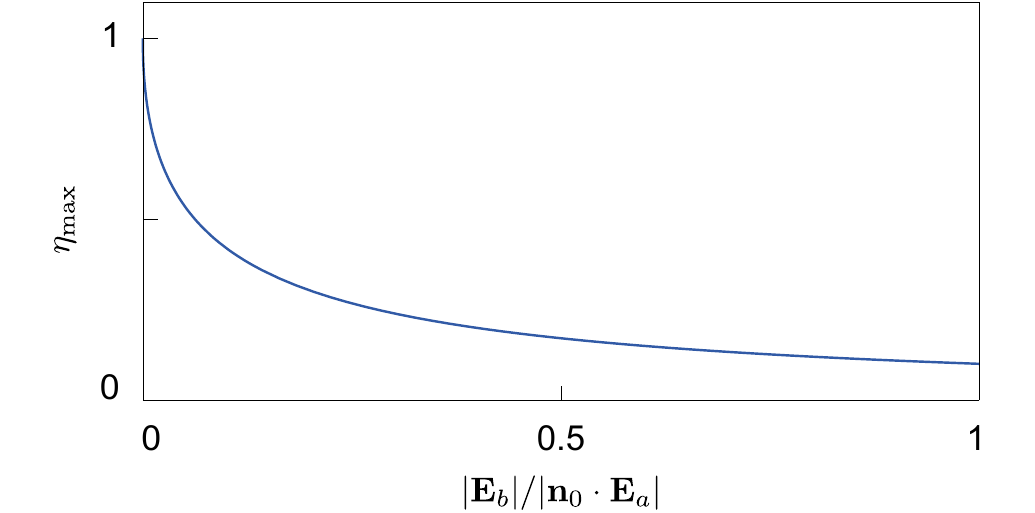}
\caption{Maximum solar cell and light amplifier efficiency $\eta_\text{max}$ as a function of $|{\bf E}_b|/|{\bf n}_0\cdot{\bf E}_{a}|$ [see Eq.(\ref{efscales}) for the definition of these electric field scales]. The maximum efficiency is achieved at $|{\bf n}_0 \cdot{\bf E}_{a}| \gg |{\bf E}_b|$ when BCD is much larger than a Drude weight.
}
\label{Fig.ESC.SCeff}
\end{figure}

To estimate quantitatively the efficiency, for simplicity we will assume a diagonal structure of the Drude weight tensor $\pmb{\mathbb D} = {\mathbb D} \,\mathbb{I}_{2\times2}$ and introduce the following notation:
\begin{equation}\label{efscales}
    {\bf E}_{a} \equiv\frac{2}{\mathbb D}\text{Re}\left[\frac{({\bf E}^*_{\omega}\times \boldsymbol{\cal D}{\bf E}_{\omega})}{1+i\omega /\Gamma}\right],~ |{\bf E}_b|^2 \equiv 2\frac{{\bf E}^*_{\omega}\cdot{\bf E}_{\omega}}{1+\omega^2/\Gamma^2},
\end{equation}
and ${\bf E}_0 = E_0{\bf n}_0$. The system can operate as a solar cell for arbitrarily small ${E}_0$ and when the sign of the circuit voltage is chosen so as to satisfy ${\bf n}_0\cdot{\bf E}_{a}>0$. Namely $\Delta W_\text{circ}$ in Eq.~(\ref{effSolar}) can becomes negative for arbitrarily small ${E}_0$. The maximum efficiency of the solar cell as a function of ${E}_0$ is obtained by finding the maximum of Eq.~(\ref{effSolar}), which occurs at: 
\begin{align}
    E_0=E_\text{Solar, max} =\sqrt{|{\bf E}_b|^2+\frac{|{\bf E}_b|^4}{|{\bf n}_0\cdot{\bf E}_a|^2}} -\frac{|{\bf E}_b|^2}{{\bf n}_0\cdot{\bf E}_a}
\end{align}
and the maximal efficiency is given by:
\begin{multline}\label{effmax}
    \eta_\text{max} =\\1 - 2 \left(\sqrt{\frac{|{\bf E}_b|^2}{|{\bf n}_0\cdot{\bf E}_a|^2}+\frac{|{\bf E}_b|^4}{|{\bf n}_0\cdot{\bf E}_a|^4}}-\frac{|{\bf E}_b|^2}{|{\bf n}_0\cdot{\bf E}_a|^2}\right).
\end{multline}

However, in order to operate as a radiation amplifier the circuit voltage direction has to be chosen to oppose the current induced by the radiation (${\bf n}_0\cdot{\bf E}_{a}<0$) and ${E}_0$ needs to overcome a threshold, given by:
\begin{equation}\label{treshholdamp}
   |{\bf E}_{\text{threshold}}| = \frac{|{\bf{E}}_b|^2}{|{\bf n}_0\cdot{\bf{E}}_a|}.
\end{equation}
The maximum of Eq.(\ref{effAmp}) as a function of $E_0$ can be found in a similar fashion (see S.I.A for details), and despite differences between requirements for the regimes, the optimal efficiency of the light amplifier is also described by Eq.~(\ref{effmax}). Notice that interestingly, in the limit $|{\bf n}_0\cdot{\bf E}_a| \gg |{\bf E}_b|$ efficiency of both devices approaches 100\% (see FIG.~\ref{Fig.ESC.SCeff}) and the threshold to reach the amplification regime given by Eq.(\ref{treshholdamp}) becomes arbitrarily small and therefore within the expected validity of the perturbative description~\footnote{We emphasize that even though it can be arbitrarily small, $\Gamma$ is always viewed as finite. This is strictly needed in order to have a well defined steady state and for quantities such as the Joule heating to remain finite.}.

The optimization that we just discussed focused on maximizing the efficiency, but in general this is not equivalent to maximizing the total delivered power [namely the maximum of the numerators of the expressions for $\eta$ in Eqs.~(\ref{effSolar}, \ref{effAmp})], which might be more relevant for practical applications. The maximum of delivered power in the solar cell regime occurs at applied voltage $E_0 = |{\bf n}_0\cdot{\bf{E}}_a|/2$, and is given by:
\begin{equation}
    \frac{\Delta W_\text{circ,max}}{T} = \frac{\mathbb D}{\Gamma}\frac{|{\bf n}_0\cdot{\bf E}_a|^2}{4},
\end{equation}
which grows with the radiation intensity and is proportional to a length of the vector $\mathrm{Im}\left[ {\bf E}^*_{\omega}\times\boldsymbol{\cal D} {\bf E}_{\omega}\right]$ for ($\Gamma\ll \omega$). On the other hand the delivered power in the light amplifying regime has no maximum within the third order of the perturbation theory and increases linearly with increasing ${E}_0$~\footnote{The optimal value will be controlled by higher order non-linear processes.}:
\begin{equation}
    \frac{\Delta W_\text{rad}}{T} =\frac{\mathbb D}{\Gamma}\left({\bf E}_0\cdot{\bf E}_a+|{\bf E}_b|^2\right).
\end{equation}

\section{Summary and discussion}

Contrary to previous claims~\cite{belinicher1986transient,ivchenko1988magneto}, we have demonstrated that it is possible for certain bulk rectification effects to induce a non-zero rectified electric current in metals when the frequency of the radiation resides within the optical gap of the material even in the limit of small relaxation rates, and shown that this is consistent with the laws of thermodynamics. We have accomplished this by using a fully microscopic description of the metallic electronic system coupled to a fermionic heat bath, and shown that this description reduces to a simpler Boltzmann single-band description within the relaxation time approximation in the limit $\Gamma \ll \omega \ll \Delta_0$, where $\Gamma$ and $\Delta_0$ are the relaxation rate and the optical gap for inter-band transitions respectively.

By considering the electronic system subjected to the simultaneous presence of a DC electric field (e.g. arising from an external circuit) and an oscillating electric field (e.g. arising from the radiation), we have shown that generically these in-gap rectification processes are irreversible and accompanied by a non-zero exchange of heat with the bath, characterized by the tensor $\pmb{\mathbb K}(\omega)$ from Eq.~(\ref{DeltaW-2nd-3rd}). We have seen that while always present, the DC Joule heating effect alone is not enough to guarantee the positivity of the net entropy production at arbitrarily small DC electric fields. Namely, in addition to the ubiquitous Joule heating, it is strictly necessary that these irreversible in-gap rectification processes [those with $\pmb{\mathbb K}(\omega) \neq 0$] in metals are accompanied by a small but finite absorption of radiation in order to guarantee the positivity of the net entropy production and abide by the second law of thermodynamics, in contrast to recent claims~\cite{onishi2022photovoltaic}. These small absorption of radiation can be provided by the tails of the Drude peak or the tails of the interband absorption at the corresponding frequency $\omega$ of the oscillating electric field that exist at small but finite relaxation rate $\Gamma$. 

We have shown, however, that the intra-band non-linear Hall effect arising from the Berry curvature dipole is special in the sense that it can be regarded as non-dissipative and reversible effect, whereby the electronic system acts as a perfect and reversible conveyor of the energy of radiation onto that energy of the circuit, and thus we have dubbed it a ``photovoltaic demon''. This allows the electronic system to operate either as a highly efficient solar cell or alternatively as an amplifier of circularly polarized light. We caution that the ``solar cell'' mode of operation requires that the radiation has some circular polarization, and therefore it is hard to imagine that this could be technologically relevant as a traditional solar cell, since sunlight is random and has no net degree of polarization. However, interestingly the amount of light absorption can be tuned with an additional DC electric field (and vanishes when this field is zero), and therefore this principle could be technologically relevant for detection and for electrical control of the transparency of circularly polarized light. On the other hand, the mode of operation in which the electronic system behaves as an amplifier of circularly polarized light holds an interesting promise as an amplifier of circularly polarized light, specially in the range of infrared frequencies.

During the completion of this work, Ref.~\cite{onishi2022photovoltaic} with some overlapping discussion on the possibility of in-gap rectification appeared, as well as Ref.~\cite{rappoport2022engineering} with a proposal for using the BCD effect for optoelectronic devices with optical gain that has some connection with our proposal of the BCD as a light amplifier. Some of our results had been preliminarily reported in~\cite{sodemann2022novel}.

\vspace{3mm}
\begin{acknowledgments}
We would like to thank Elio K\"onig, Adolfo Grushin and Fernando de Juan for estimulating discussions. I. S. would like to specially thank Urmimala Dey who performed several unpublished calculations that served as motivation for this project. J. C. W. S. acknowledges support from the Ministry of Education, Singapore under its MOE AcRF Tier 3 Grant No. MOE2018-T3-1-002.
\end{acknowledgments}

\bibliography{BCD-ingap-rectificationBIB}

\clearpage

\appendix

\renewcommand{\theequation}{S-\arabic{equation}}
\renewcommand{\thefigure}{S-\arabic{figure}}
\renewcommand{\thetable}{S-\Roman{table}}
\makeatletter
\renewcommand\@biblabel[1]{S#1.}
\setcounter{equation}{0}
\setcounter{figure}{0}

\onecolumngrid

\section{Boltzmann equation, perturbation theory, solar cell and a light amplifier}\label{sec:BZeqsupp}
In this section we will derive corrections to the electron distribution function and electric current in the presence of electric field, starting from Boltzmann equation:
\begin{equation}
    \partial_t f + {\bf E}(t) \cdot \partial_{\bf k} f = \Gamma(f_0 - f),\qquad
    {\bf E}(t)= {\bf E}_0 +{\bf E}_\omega e^{i \omega t} +  {\bf E}_{-\omega}^{} e^{-i \omega t},
\end{equation}
\begin{equation}
    f = \sum_{n=0} ^{\infty} f_n(t), \quad \text{where} \quad f_0 = f_{F-D},\quad \text{and}\quad f_n \sim |{\bf E}|^n,\quad {\bf E}_{\omega}^* = {\bf E}_{-\omega}^{}. 
\end{equation}
$f_{F-D}$ stands for a Fermi-Dirac distribution. Iterative solution of equations above brings us to the following conclusion:
\begin{gather}
    f_1(t) = f_1(0)+f_1(\omega)e^{i\omega t}+f_1(-\omega)e^{-i\omega t},\\
    f_2(t) = f_2(0)+f_2(\omega)e^{i\omega t}+f_2(2\omega)e^{i2\omega t}+f_2(-\omega)e^{-i\omega t}+f_2(-2\omega)e^{-2i\omega t},
\end{gather}
where $f(-\omega) = f(\omega)^*$ and:
\begin{gather}
    f_1(0) = -\frac{1}{\Gamma} {\bf E}_0\cdot \partial_{\bf{k}} f_0,\quad f_1(\omega) = -\frac{1}{\Gamma+i\omega} {\bf E}_\omega\cdot \partial_{\bf{k}} f_0,\\
    f_2(0) = -\frac{1}{\Gamma} \left({\bf E}_0\cdot \partial_{\bf k} f_1(0)+{\bf E}_\omega\cdot \partial_{\bf k} f_1(-\omega)+{\bf E}_{-\omega}\cdot \partial_{\bf k} f_1(\omega)\right),\\ f_2(\omega) = -\frac{1}{\Gamma+i\omega}{\bf E}_{\omega}\cdot\partial_{\bf k}f_1(0)-\frac{1}{\Gamma+i\omega}{\bf E}_{0}\cdot\partial_{\bf k}f_1(\omega),\\
    f_2(2\omega) = -\frac{1}{\Gamma+2i \omega} {\bf E}_{\omega}\cdot\partial_{\bf k} f_1 (\omega),\quad \text{where}\quad {\bf a}\cdot\partial_{\bf k}\equiv \sum_i a^i \frac{\partial}{\partial k^i},
\end{gather}
which allows us to compute the electric current response to electric field. In the first order we obtain :
\begin{gather}
   {\bf j}^{(1)}(t) ={\bf j}_1(0) + {\bf j}_1(\omega)e^{i\omega t}+ {\bf j}^{(1)}(-\omega)e^{-i\omega t},\\
   {\bf j}^{(1)}(0) = f_1(0)\partial_{\bf k}\epsilon + f_0 \left[{\bf \Omega}\times {\bf E}_0\right],\\
   {\bf j}^{(1)}(\pm\omega) = f_1(\pm\omega)\partial_{\bf k}\epsilon + f_0 \left[{\bf \Omega}\times {\bf E}_{\pm\omega}\right],
\end{gather}
and in the second order:
\begin{gather}
    {\bf j}^{(2)}(t) ={\bf j}^{(2)}(0) + \left({\bf j}^{(2)}(\omega)e^{i\omega t}+ {\bf j}^{(2)}(2\omega)e^{2i\omega t} + \text{c.c.}\right),\\
    {\bf j}^{(2)}(0) = f_2(0)\partial_{\bf k}\epsilon + f_1(0) \left[{\bf \Omega}\times {\bf E}_0\right]+ f_1(\omega) \left[{\bf \Omega}\times {\bf E}_{-\omega}\right]+f_1(-\omega) \left[{\bf \Omega}\times {\bf E}_{\omega}\right],\\
    {\bf j}^{(2)}(\omega) = f_2(\omega)\partial_{\bf k}\epsilon + f_1(0) \left[{\bf \Omega}\times {\bf E}_\omega\right]+ f_1(\omega) \left[{\bf \Omega}\times {\bf E}_{0}\right],\\
    {\bf j}^{(2)}(2\omega) = f_2(2\omega)\partial_{\bf k}\epsilon + f_1(\omega) \left[{\bf \Omega}\times {\bf E}_\omega\right].
\end{gather}

Before moving to the computation of the work done by a system, we want to emphasise that the total electric field has two physically different components: a DC component that represents a circuit voltage and an AC component that represents incoming radiation. These two components do a work separately and thus we split their contributions accordingly $\Delta  W = \Delta W_\text{circ} + \Delta W_\text{rad}$, where $\Delta W_\text{rad}$ is a work/power performed by the radiation,
the $\Delta W_\text{circ}$ work done by the circuit, and they are given by: 
\begin{equation}
   \Delta W_\text{circ} = \int_{t_i}^{t_f}  {\bf j}(t)\cdot {\bf E}_0\text{d} t,\qquad \Delta W_\text{rad} = \int_{t_i}^{t_f}  {\bf j}(t)\cdot {\bf E}_\omega (t)\text{d} t. 
\end{equation}  
The sum of two quantities above has to be non negative, which brings us to three possible regimes: 
\begin{enumerate}
  \item $\Delta W_\text{circ}\geq 0$, $\Delta W_\text{rad}\geq 0$. In this regime system absorbs energy from all the incoming radiation.
  \item $\Delta W_\text{circ}\geq 0$, $\Delta W_\text{rad}\leq 0$, $|W_\text{circ}|\geq |W_\text{rad}|$. In this regime system takes energy from a radiation and delivers part of it to a circuit. Which is a solar cell.
  \item $\Delta W_\text{circ}\leq 0$, $\Delta W_\text{rad}\geq 0$, $|W_\text{circ}|\leq |W_\text{rad}|$. In this regime system takes energy from a circuit and delivers part of it into a radiation. Which is a light amplifier.
\end{enumerate}

First, let us consider the Berry dipole current ${\bf j}_\text{BD}(t)$ and it's averaged power, which we separated into the absorbed power $\Delta W _\text{rad,BD} = {\bf j}_\text{BD}(\omega) \cdot {\bf E}_\omega^* + {\bf j}_\text{BD}(-\omega) \cdot {\bf E}_\omega $ done by incident radiation with oscillating field, and the delivered power $\Delta W _\text{circ,BD} = {\bf j}_\text{BD}(0) \cdot {\bf E}_0$ done on to the electric circuit by the constant electric field.
Using solution of the Boltzmann equation from above, we obtain:
\begin{gather}
    {\bf j}_\text{BD}(0) = \int_{\bf k}  \big( f_1(0) \left[{\bf \Omega} \times {\bf E}_0\right] + f_1(\omega) [{\bf \Omega} \times {\bf E}_{-\omega}] + f_1(-\omega) [{\bf \Omega} \times {\bf E}_\omega] \big),\\
    {\bf j}_\text{BD}(\omega) = \int_{\bf k}  \big( f_1(0) [{\bf \Omega} \times {\bf E}_\omega] + f_1(\omega) [{\bf \Omega} \times {\bf E}_0 ]\big).
\end{gather}
With these obtained, the Berry dipole related absorbed power and delivered power are
\begin{align}
& \Delta W _\text{rad,BCD} = \int_{\bf k} \big\{ f_1(\omega) ([{\bf \Omega} \times {\bf E}_{0}] \cdot {\bf E}_{-\omega}) + f_1({-\omega}) ([{\bf \Omega} \times {\bf E}_0] \cdot {\bf E}_{\omega}) \big\}=-2 {\bf E}_0\cdot \text{Re}\left[\frac{\boldsymbol{\cal D} {\bf E}_{\omega}\times{\bf E}_{-\omega}}{\Gamma+i\omega} \right] , \label{PaBD}
\\
& \Delta W _\text{circ,BCD} = \int_{\bf k} \big\{ f_1(\omega)([ {\bf \Omega} \times {\bf E}_{-\omega}] \cdot {\bf E}_0) + f_1({-\omega}) ([{\bf \Omega} \times {\bf E}_{\omega}] \cdot {\bf E}_0) \big\}=2 {\bf E}_0\cdot \text{Re}\left[\frac{\boldsymbol{\cal D} {\bf E}_{\omega}\times{\bf E}_{-\omega}}{\Gamma+i\omega} \right], \label{PdBD}
\end{align}
where ${\cal D}_{a b} = \int_{\bf k} f_0 \partial_a \Omega_b $ is the Berry dipole and $[\boldsymbol{\cal D} {\bf E}]_a = \sum_b {\cal D}_{a b} E_b$ is a matrix-vector multiplication. These two powers exactly cancel each other (${\bf a}\cdot [{\bf b}\times {\bf c}]=-{\bf c}\cdot [{\bf b}\times {\bf a}]$), $\Delta W _\text{rad,BD} + \Delta W _\text{circ,BD} = 0$, which agrees with the above general analysis that total power from the Berry dipole related current vanishes at any order of the perturbation theory.

Now, let us write the total energy delivered and absorbed up to the second order of the perturbation theory for current (third order for power), which after slight simplifications, can be written as:
\begin{align}
& \Delta W _\text{rad} =\frac{2\Gamma}{\Gamma^2+\omega^2} {\bf E}_{-\omega}\pmb{\mathbb D}{\bf E}_\omega+\frac{4\Gamma^2}{(\Gamma^2+\omega^2)^2}{\bf E}_0\pmb{\mathbb J} {\bf E}_\omega{\bf E}_{-\omega}-2 {\bf E}_0\cdot \text{Re}\left[\frac{\boldsymbol{\cal D} {\bf E}_{\omega}\times{\bf E}_{-\omega}}{\Gamma+i\omega} \right] , \label{PaBDsup}
\\
& \Delta W _\text{circ} =\frac{1}{\Gamma} {\bf E}_0\pmb{\mathbb D}{\bf E}_0+\frac{1}{\Gamma^2}{\bf E}_0\pmb{\mathbb J} {\bf E}_0{\bf E}_0+\frac{1}{\Gamma^2+\omega^2}{\bf E}_0\pmb{\mathbb J} {\bf E}_\omega{\bf E}_{-\omega}+2 {\bf E}_0\cdot \text{Re}\left[\frac{\boldsymbol{\cal D} {\bf E}_{\omega}\times{\bf E}_{-\omega}}{\Gamma+i\omega} \right], \label{PdBDsup}
\end{align}
where ${\bf A} \pmb{\mathbb D} {\bf B} = \sum_{a b} A_a {\mathbb D}_{ab} B_b$, ${\bf A} \pmb{\mathbb J} {\bf B} {\bf C}  = \sum_{abc} A_a {\mathbb J}_{abc} B_b C_c$ and:
\begin{equation}
    {\mathbb D}_{ab} = \int_{\bf k} f_0 \partial_a \partial_b \epsilon, \qquad {\mathbb J}_{abc} = \int_{\bf k} f_0 \partial_a \partial_b \partial_c \epsilon.
\end{equation}
are the Drude weight and Jerk tensors. 
We see that delivered Eq.(\ref{PaBDsup}) and absorbed Eq.(\ref{PdBDsup}) powers are sensitive to a sign of a circuit voltage. If the Drude weight is negligible electro-optic effect is dominant, which enables an unexpected regime of powering the radiation from a circuit. Additionally, if the circuit voltage direction is switched the system transits into a solar cell regime. 

We note that the requirement:
\begin{equation}\label{PTrestr-S}
    \Delta W _\text{rad}+\Delta W _\text{circ}=\frac{2\Gamma}{\Gamma^2+\omega^2} {\bf E}_{-\omega}\pmb{\mathbb D}{\bf E}_\omega+\frac{1}{\Gamma} {\bf E}_0\pmb{\mathbb D}{\bf E}_0+\frac{6\Gamma^2+2\omega^2}{(\Gamma^2+\omega^2)^2}{\bf E}_0\pmb{\mathbb J} {\bf E}_\omega{\bf E}_{-\omega}+\frac{1}{\Gamma^2}{\bf E}_0\pmb{\mathbb J} {\bf E}_0{\bf E}_0\geq 0 
\end{equation}
set's a limit of perturbation theory validity. We see that BCD current is dissipationless and in not present in Eq.(\ref{PTrestr-S}) whereas Jerk current is dissipative. It is important to notice that both effects are finite in an optical gap even in a clean limit $\Gamma \rightarrow0$. Interestingly, in a limit $\pmb{\mathbb J}\rightarrow 0$ the restriction Eq.(\ref{PTrestr-S}) is automatically satisfied due to a positivity of a Drude weight. Yet, in general, this requirement may not be satisfied for arbitrary value of ${\bf E}_0$. For example in a limit $ \omega \gg \Gamma$ we obtain:
\begin{equation}
    \Gamma{\bf E}_0\pmb{\mathbb D}{\bf E}_0+{\bf E}_0\pmb{\mathbb J} {\bf E}_0{\bf E}_0\geq 0, 
\end{equation}
which defines limits for a perturbation theory validity.

In the remaining part of the section, we want to demonstrate how to use our theory to optimise the performance of a system as a solar cell or a light amplified. First, assuming that $\Delta W _\text{rad}<0$, $\Delta W _\text{circ}>0$ which means that system operates as a solar cell and time-reversal symmetry ($\pmb{\mathbb J} = 0$) we want to analyze an efficiency of the system:
\begin{equation}
    \eta_S=-\frac{\Delta W_{\text{circ}}}{\Delta W_{\text{rad}}}= \frac{ {\bf E}_0\cdot \text{Re}\left[\frac{{\bf E}_{-\omega}\times\boldsymbol{\cal D} {\bf E}_{\omega}}{1+i\omega/\Gamma} \right]-\frac{1}{2} {\bf E}_0\pmb{\mathbb D}{\bf E}_0}{{\bf E}_0\cdot \text{Re}\left[\frac{{\bf E}_{-\omega}\times\boldsymbol{\cal D} {\bf E}_{\omega}}{1+i\omega/\Gamma} \right]+\frac{{\bf E}_{-\omega}\pmb{\mathbb D}{\bf E}_\omega}{1+\omega^2/\Gamma^2} }.
\end{equation}
To simplify the further analysis we also assume that Drude weight is a diagonal tensor $\pmb{\mathbb D} = {\mathbb D} \, \mathbb{I}_{2\times2}$, allowing us to rewrite the efficiency in a simplified form:
\begin{equation}
    \eta_\text{Solar} = \frac{{\bf E}_0\cdot{\bf E}_a - |{\bf E}_0|^2}{{\bf E}_0\cdot{\bf E}_a +|{\bf E}_b|^2},\quad \text{where}\quad {\bf E}_0 = E_0{\bf n}_0, \quad {\bf E}_a =\frac{2}{\mathbb D}\text{Re}\left[\frac{{\bf E}_{-\omega}\times \boldsymbol{\cal D}{\bf E}_{\omega}}{1+i\omega /\Gamma}\right],\quad |{\bf E}_b|^2 = 2\frac{{\bf E}_{-\omega}\cdot{\bf E}_{\omega}}{1+\omega^2/\Gamma^2},
\end{equation}
which in a limit ${\cal D} \gg {\mathbb D}$ can approach 1 (it is transparently seen if we also assume diagonal structure of a Berry dipole, however it is not needed in general). We emphasize that limit of ultimate efficiency is achieved when Drude weight is negligible. This regime is physically distinct from a clean limit, where Joel heating becomes immense for arbitrary small value of a circuit voltage. In a clean limit BCD mechanism is possible, however, one can not use it to power a solar cell. 

Next, we study optimization of the device performance by tuning the applied voltage. It can be shown, that maximum efficiency of a solar cell (which is possible for ${\bf n}_0\cdot{\bf E}_a>0$ and $E_0<{\bf n}_0\cdot{\bf E}_a$) is expected at the following voltage:
\begin{equation}
    E_0 =\sqrt{|{\bf E}_b|^2+\frac{|{\bf E}_b|^4}{|{\bf n}_0\cdot{\bf E}_a|^2}} -\frac{|{\bf E}_b|^2}{{\bf n}_0\cdot{\bf E}_a} \quad \rightarrow \quad \text{max}[\eta_S] =1 - 2 \left(\sqrt{\frac{|{\bf E}_b|^2}{|{\bf n}_0\cdot{\bf E}_a|^2}+\frac{|{\bf E}_b|^4}{|{\bf n}_0\cdot{\bf E}_a|^4}}-\frac{|{\bf E}_b|^2}{|{\bf n}_0\cdot{\bf E}_a|^2}\right).
\end{equation}
Note that maximization of delivered power occurs at a different voltage $\max[ \Delta W_{\text{circ}}]\Rightarrow E_0 = {\bf n}_0\cdot{\bf E}_a/2 \rightarrow W_\text{circ,max} = |{\bf n}_0\cdot{\bf E}_a|^2/4$.

Similar analysis can be done for maximization of an efficiency of light amplifier with time-reversal symmetry (which is possible for ${\bf n}_0\cdot{\bf E}_a<0$ and $E_0>|{\bf E}_b|^2/|{\bf n}_0\cdot{\bf E}_a|$). In this case we have $\Delta W _\text{rad}<0$ and obtain:
\begin{equation}
    \eta_\text{Amp}=-\frac{\Delta W_{\text{rad}}}{\Delta W_{\text{circ}}}=\frac{{\bf E}_0\cdot{\bf E}_a +|{\bf E}_b|^2}{{\bf E}_0\cdot{\bf E}_a - |{\bf E}_0|^2}.
\end{equation}
Which is maximised at the following electric field with the consequent maximum efficiency:
\begin{equation}
    E_0 =\sqrt{|{\bf E}_b|^2+\frac{|{\bf E}_b|^4}{|{\bf n}_0\cdot{\bf E}_a|^2}} +\frac{|{\bf E}_b|^2}{|{\bf n}_0\cdot{\bf E}_a|},\quad \rightarrow\quad \text{max}[\eta_\text{Amp}]=1 - 2 \left(\sqrt{\frac{|{\bf E}_b|^2}{|{\bf n}_0\cdot{\bf E}_a|^2}+\frac{|{\bf E}_b|^4}{|{\bf n}_0\cdot{\bf E}_a|^4}}-\frac{|{\bf E}_b|^2}{|{\bf n}_0\cdot{\bf E}_a|^2}\right).
\end{equation}
Interestingly enough, the optimal efficiency of the light amplifier is the same as the solar cell's, where ultimate efficiency is achieved when the Drude weight is negligible compared to a Berry dipole. Yet, in amplifying regime, amplifying power has no optimal regime. The amplifying power linearly increases with the electric field magnitude.
\end{document}